\documentstyle[stwol,epsfig]{article}
\input{psfig}
\input{epsf}
\def\be{\begin{equation}}
\def\eea{\end{eqnarray}}

\def\D0{D$\!$\O}

\def\pdfs   {parton distribution functions}
\def\vetmis {\mbox{${\hbox{${\bf E}$\kern-0.6em\lower-.1ex\hbox{/}}}_T$ }}
\def\etmisv {\mbox{${\hbox{${\vec E}$\kern-0.6em\lower-.1ex\hbox{/}}}_T$ }}
\def\etmiss {\mbox{${\hbox{$E$\kern-0.6em\lower-.1ex\hbox{/}}}_T$ }}
\def\etmis  {\mbox{${\hbox{$E$\kern-0.6em\lower-.1ex\hbox{/}}}_T$ }}
\def\ptmiss {\mbox{${\hbox{$p$\kern-0.6em\lower-.1ex\hbox{/}}}_T$ }}
\def\ptmis  {\mbox{${\hbox{$p$\kern-0.6em\lower-.1ex\hbox{/}}}_T$ }}

\def\pbarp  {\mbox{$\overline{p}p$ }}

\def\Zeemis 
    {\mbox{${\hbox{$Z\rightarrow ee$\kern-0.4em\lower-.1ex\hbox{/}}}$ }}
\def\pz{\phantom{0}}
\def\pzz{\phantom{00}}
\def\ifmath#1{\relax\ifmmode #1\else $#1$\fi}%
\def\GeV{\ifmmode {\mathrm{ Ge\kern -0.1em V}}\else
                   \textrm{Ge\kern -0.1em V}\fi}%
\def\MeV{\ifmmode {\mathrm{ Me\kern -0.1em V}}\else
                   \textrm{Me\kern -0.1em V}\fi}%
\def\keV{\ifmmode {\mathrm{ ke\kern -0.1em V}}\else
                   \textrm{ke\kern -0.1em V}\fi}%
\def\eV{\ifmmode  {\mathrm{ e\kern -0.1em V}}\else
                   \textrm{e\kern -0.1em V}\fi}%
\def\GeVcc{\ifmmode {\mathrm{ \GeV/c^2}}\else
                   \textrm{Ge\kern -0.1em V/c$^2$}\fi}%

\newcommand{\mch}      {\multicolumn {2} {|c|}}
\newcommand{\SM}        {\mbox{Standard Model}}
\newcommand{\nb}        {\rm{nb}}

\newcommand{\qq}        {\mathrm{q}\overline{\mathrm{q}}}
\newcommand{\MZ}        {m_{\mathrm{Z}}}
\newcommand{\MW}        {m_{\mathrm{W}}}
\newcommand{\MH}        {m_{\mathrm{H}}}
\newcommand{\Mt}        {m_{\mathrm{t}}}
\newcommand{\shad}      {\sigma_{\mathrm{h}}^{0}}
\newcommand{\RZ}        {R_{\ell}}

\newcommand{\AFB}       {A_{\mathrm{FB}}}
\newcommand{\Afbpole}   {A^{0,\,\ell}_{\rm {FB}}}
\newcommand{\AFBpole}   {A^{0,\,\ell}_{\rm {FB}}}
\newcommand{\Afbzl}     {A^{0,\,\ell}_{\rm {FB}}}
\newcommand{\Afbze}     {A^{0,\,{\rm e}}_{\rm {FB}}}

\newcommand{\Afbzm}     {A^{0,\,\mu}_{\rm {FB}}}
\newcommand{\Afbzt}     {A^{0,\,\tau}_{\rm {FB}}}

\newcommand{\Afbzb}     {A^{0,\,b}_{\rm {FB}}}
\newcommand{\Afbzc}     {A^{0,\,c}_{\rm {FB}}}
\newcommand{\ALR}       {\mbox{$A_{\rm {LR}}$}}
\newcommand{\ALRpole}   {\mbox{$A_{\rm {LR}}^{0}$}}

\newcommand{\AFBtau}    {\mbox{$A_{\rm {FB}}^{\cal P_{\tau}}$}}

\newcommand{\avQfb}     {\langle \mbox{Q}_{\mbox{FB}} \rangle}
\newcommand{\ptau}      {\mbox{$\cal P_{\tau}$}}
\newcommand{\cAe}       {\mbox{$\cal A_{\rm e}$}}
\newcommand{\cAt}       {\mbox{$\cal A_{\tau}$}}
\newcommand{\cAf}       {\mbox{$\cal A_{\rm f}$}}

\newcommand{\cAl}       {\mbox{$\cal A_{\ell}$}}
\newcommand{\cAb}       {\mbox{$\cal A_{\rm b}$}}
\newcommand{\cAc}       {\mbox{$\cal A_{\rm c}$}}

\newcommand{\GZ}        {\Gamma_{\mathrm{Z}}}
\newcommand{\Gll}       {\Gamma_{\ell\ell}}

\newcommand{\Ginv}      {\Gamma_{\mathrm{inv}}}
\newcommand{\Ghad}      {\Gamma_{\mathrm{had}}}
\newcommand{\Gnu}       {\Gamma_{\nu\nu}}
\newcommand{\Gbb}       {\ifmath{\Gamma_{\mathrm{b\bar{b}}}}}
\newcommand{\Gcc}       {\ifmath{\Gamma_{\mathrm{c\bar{c}}}}}

\newcommand{\Rl}        {R_{\ell}}
\newcommand{\Rb}        {\ifmath{R_{\mathrm{b}}}}

\newcommand{\Rc}        {\ifmath{R_{\mathrm{c}}}}

\newcommand{\swsq}      {1-\MW^2/\MZ^2}

\newcommand{\swsqeffl}  {\sin^2\!\theta_{\rm{eff}}^{\rm {lept}}}

\newcommand{\gahatf}    {g_{A}^{f}}
\newcommand{\gvhatf}    {g_{V}^{f}}
\newcommand{\gahatl}    {g_{A}^{\ell}}
\newcommand{\gvhatl}    {g_{V}^{\ell}}
\newcommand{\gahatn}    {g_{A}^{\nu}}
\newcommand{\gvhatn}    {g_{V}^{\nu}}
\newcommand{\ghatn}     {g_{\nu}}
\newcommand{\ff}        {\mbox{$f\overline{f}$ }}
\newcommand{\ee}        {\mbox{$e^+e^-$}}
\newcommand{\bb}        {\mbox{$b\overline b$}}

\newcommand{\alphas}    {\alpha_s}
\newcommand{\alfmz}     {\alphas(\MZ^2)}
\newcommand{\effb}      {\varepsilon_{b}}
\newcommand{\effc}      {\varepsilon_{c}}
\newcommand{\effuds}    {\varepsilon_{uds}}

\newcommand{\Dstar}     {{\rm D}^{* }}
\newcommand{\Dstarp}    {{\rm D}^{*+}}

\newcommand{\Dzero}     {{\rm D}^0}
\newcommand{\Dplus}     {{\rm D}^+}

\newcommand{\Ds}        {{\rm{D_s}}}
\newcommand{\Lc}        {{\rm{\Lambda_c}}}
\newcommand{\RcPcX}     {\mbox{${\Gcc \over \Ghad} \cdot 
                                {\rm P}( c \rightarrow {\rm X}_c) \cdot 
                                 BR_{{\rm X}_c} $}}
\newcommand{\PcDst}     {\mbox{$\mathrm{P( c \rightarrow D^{*+}) \times
                        BR( D^{*+} \rightarrow \pi^+ D^0 )}$}}
\newcommand{\blX}       {\mbox{$b{\mathrm \rightarrow \ell X}$}}
\newcommand{\clX}       {\mbox{$c{\mathrm \rightarrow \ell X}$}}
\newcommand{\BrblX}     {BR(\blX )}
\newcommand{\BrclX}     {BR(\clX)}
\newcommand{\mcc}[1]    {\multicolumn{1}{c|}{#1}}
\begin{document}

\begin{titlepage} 
\rightline{\vbox{\halign{&#\hfil\cr
&Fermilab-Conf-96/354\cr
&November 1996\cr}}}
\vspace{1in} 

\renewcommand{\thefootnote}{\fnsymbol{footnote}} 
\setcounter{footnote}{2}

\begin{center}

{\Large\bf
    Precision Electroweak Measurements\footnote{\normalsize 
    Invited talk given at the 
    {\it  Meeting of the Division of Particles and
      Fields}, 
    Minneapolis, August 10 -- 15, 1996.}
}

\vskip 2.0cm
\normalsize 
{\large Marcel Demarteau} 
\vskip .3cm
Fermilab \\
Batavia, IL 60510, USA\\
\vskip 3cm

\end{center}

\begin{abstract} 

Recent electroweak precision measurements from $\ee$ and $\pbarp$
colliders are presented. Some emphasis is placed on the recent
developments in the heavy flavor sector. The measurements are compared to 
predictions from the Standard Model of electroweak interactions. 
All results are found to be consistent with the Standard Model. 
The indirect
constraint on the top quark mass from all measurements is in
excellent agreement with the direct $\Mt$ measurements. 
Using the world's electroweak data in conjunction with the current 
measurement of the top quark mass, the constraints on the Higgs mass are
discussed. 

\end{abstract} 

\setcounter{footnote}{0}
\renewcommand{\thefootnote}{\arabic{footnote}} \end{titlepage}

\title{Precision Electroweak Measurements }
\author{ Marcel Demarteau }
\address{\it Fermilab, P.O. Box 500, 
            Batavia, IL 60510, USA}

%
%
%

\twocolumn[\maketitle\abstracts{
Recent electroweak precision measurements from $\ee$ and $\pbarp$
colliders are presented. Some emphasis is placed on the recent
developments in the heavy flavor sector. The measurements are compared to 
predictions from the Standard Model of electroweak interactions. 
All results are found to be consistent with the Standard Model. 
The indirect
constraint on the top quark mass from all measurements is in
excellent agreement with the direct $\Mt$ measurements. 
Using the world's electroweak data in conjunction with the current 
measurement of the top quark mass, the constraints on the Higgs mass are
discussed. 
}]

\section{Introduction}

Radiative corrections in the standard model of electroweak interactions
($\SM$) have taken a very prominent position in today's description of 
experimental results. 
Perhaps the most compelling reason for this state of affairs is that 
the experimental results have reached a level of precision which require 
a comparison with theory beyond the Born calculations, which the $\SM$ 
is able to provide. 
If loop calculations are needed for the calculation of physics
observables, the measurements show sensitivity to the masses and
couplings of the particles propagating in the loops. 
The experimental measurements can thus provide information about the 
particles contributing to the radiative corrections well below the 
threshold for directly producing them. 

In this summary the most recent electroweak results 
from $\ee$ and $\pbarp$ colliders will be described. 
The emphasis will be on 
electroweak results from data taken on the $Z$ resonance. 
Within the $\SM$, the description of all processes involving 
neutral currents is given in terms of 
the chiral couplings of the fermion $f$ to the $Z$ boson, 
$g_L^f$ and $g_R^f$, or more commonly in terms of the vector and
axial-vector couplings, $\gvhatf$ and $\gvhatf$: 
\begin{eqnarray*}
    \gvhatf = & (g_L^f \,+\, g_R^f) 
            = & I^f_3 \,-\, 2 \, Q_f \, \sin^2\vartheta_W \\
    \gahatf = & (g_L^f \,-\, g_R^f) = & I^f_3  \ .
\end{eqnarray*}
Here $\vartheta_W$ is the weak mixing angle, $I^f_3$ the weak isospin
component of fermion $f$ and $Q_f$ its charge.

Because the left-handed and right-handed coupling of fermions to the 
$Z$ boson are not the same, the angular distribution of the 
outgoing fermion with respect to the incoming fermion in the 
center of mass frame for the process $\ee\rightarrow\ff$
has a term linear in $\cos(\vartheta)$~\cite{angle}. 
The distribution is thus 
asymmetric and will exhibit a forward-backward asymmetry, defined as
\begin{displaymath}
    \AFB \, = \, {\sigma_F \,-\, \sigma_B
                  \over
                  \sigma_F \,+\, \sigma_B }         \ ,
\end{displaymath}
where $\sigma_F$ is the cross section for fermion production in the
forward hemisphere ($0^\circ < \vartheta < 90^\circ$) and
$\sigma_B$ the cross section for the backward hemisphere
($90^\circ < \vartheta < 180^\circ$).

Around the $Z$ pole, the photon exchange and $\gamma Z$ interference
are only small corrections to the resonance cross section. Retaining only
the resonance cross section, the forward-backward asymmetry on the pole 
is, at lowest order, given by 
\begin{displaymath}
    \AFBpole \, = \,    {3\over 4} \, {\cal A}_e \, {\cal A}_\ell \ ,
\end{displaymath} 
where the asymmetry of couplings, $\cAf$, are given by 
\begin{displaymath}
    \cAf    \,\equiv\, { {g_L^f}^2 \,-\, {g_R^f}^2  \over 
                         {g_L^f}^2 \,+\, {g_R^f}^2  } 
            \,=\,      {2\, g_V^f \, g_A^f \over 
                           {g_V^f}^2 \,+\, {g_A^f}^2 }  \ .
\end{displaymath} 
These expressions will be modified when including higher order
corrections. Weak vertex corrections and self-energy diagrams will 
introduce fermion dependent form factors which can be absorbed in the
definition of the coupling constants~\cite{hollik}. 
By introducing effective 
coupling constants the Born structure of the processes can 
to a good approximation be retained. 
Since all asymmetry measurements 
determine essentially the ratio of couplings 
${g_V^f} / {g_A^f}$ it is convenient to define an effective 
electroweak mixing angle 
\begin{displaymath}
    \swsqeffl   \,\equiv\, {1\over 4} \, 
                           \left( 1 - {{g_V^f}^2 \over {g_A^f}^2} 
                           \right) \ \ ,
\end{displaymath} 
which is well matched with the quantities measured experimentally. 
The effective electroweak mixing angle is, coincidentally, very close 
to the definition in the ${\overline {\rm MS}}$ scheme~\cite{sirlin}. 

The results presented are based on event samples of 
about $4 \cdot 10^5$ 
leptonic and $3.5 \cdot 10^6$ hadronic $Z$ decays per LEP experiment, 
complemented with $1.6 \cdot 10^5$ $Z$ decays recorded at SLC with 
a polarized electron beam. 
The \pbarp experiments CDF and \D0 each have 
approximately 60,000 leptonic $W$ and 6000 leptonic $Z$ decays 
collected during the 1992-1993 run (Run~1a) and  
the 1994-1995 run (Run~1b) combined. A fivefold increase in luminosity was
obtained in the latter run. 
All results presented are preliminary. 

In the next section results from line shape measurements will be
described. The results on the effective coupling constants from the
line shape measurements and the forward-backward asymmetries in the
leptonic sector will be summarized in section~\ref{sec:avg}. 
The section following describes the main developments in the hadronic 
sector with the emphasis on $\Rb$ and $\Rc$. 
Given the full set of measurements, 
an overall fit is performed within the framework of the $\SM$ and the 
consistency of the results verified. 
We will conclude with some recent developments.

%

\section{Line shapes and Asymmetries } 
\subsection{Line shape Measurements } 

\begin{figure}[h]
    \epsfxsize = 7.0cm
    \centerline{\epsffile{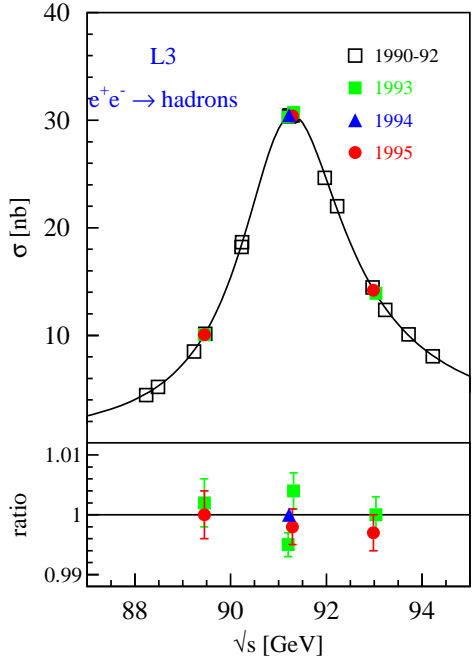}}
\caption[]{Hadronic resonant cross section as function of $\sqrt{s}$ 
           as measured by L3 (top) and the comparison with the theoretical 
           prediction (bottom).}
\label{fig:l3_xsec}
\end{figure}

In \pbarp and $\ee$ collisions final state lepton pairs are produced
through photon and $Z$ exchange. The cross section at lowest order 
is given by 
\begin{eqnarray*} 
    \sigma_{ff} &\propto & {12\pi \over M_Z^2} \, 
                           {\Gamma_{ee} \Gamma_{ff} \over \Gamma_Z^2 } \, 
                           { s \, \Gamma_Z^2      \over 
                            (s-M_Z^2 )^2 \,+\, s^2 \Gamma_Z^2 / M_Z^2 } 
                           \\
                & &  +\,  ``\gamma Z"     \,+\,   
                          ``\gamma "
\end{eqnarray*} 
consisting of the $Z$ resonance cross section, 
the QED annihilation term (\lq\lq$\gamma$\rq\rq) and the 
$\gamma Z$ interference term (\lq\lq$\gamma Z$\rq\rq). 
At proton colliders the resonance cross section for $W$ and $Z$
production is used to indirectly 
determine the width of the $W$ boson through the ratio of the $W$ and
$Z$ production cross sections. At $\ee$ colliders the measurement of the
resonance line shape is used to extract $\MZ$ and $\GZ$. 
Figure~\ref{fig:l3_xsec} shows the hadronic resonant cross section 
as measured by the L3 experiment. 
From the hadronic decays of the $Z$ boson 
the hadronic pole cross section, 
$\shad \equiv {12\pi \over M_Z^2} \, 
{\Gamma_{ee} \Gamma_{\rm had} \over \Gamma_Z^2 }$
is determined. 
From the leptonic decays the ratio of
the partial hadronic and leptonic widths, 
$R_\ell \equiv {\Ghad \over \Gll }$, is derived. 
This particular choice of variables, $\MZ, \GZ, \shad$ and $\RZ$, 
is motivated by the desire to minimize the correlation among 
the variables and to minimize any model dependence. 
One of the main challenges of these measurements is to control the
systematic uncertainties and keep them at the same level as the
statistical uncertainties. 
Since the measurements of these quantities entail both an absolute cross
section measurement and an absolute mass determination, the luminosity
and energy calibration are crucial. 

The LEP experiments all measure the luminosity with small angle silicon 
based calorimeters with good spatial resolution counting Bhabha events. 
At small scattering angles $\vartheta$, 
the cross section for Bhabha scattering shows a $\vartheta^{-3}$ 
dependence. For the luminosity measurement a very precise knowledge of
the edges of the acceptance is required. An accuracy of 10~$\mu$m is 
currently achieved, resulting in an uncertainty of 
$\delta({\cal L}) \approx  (0.07 - 0.15)$\%, surpassing the theoretical
uncertainty~\cite{lumi_95}. 
Recent advances in the calculation of the Bhabha cross
section~\cite{bhabha} have significantly reduced the theoretical 
uncertainty on the luminosity to the level of~0.11\%, 
with a further reduction of a factor of
two anticipated in the near future. 

\begin{figure}[h]
    \epsfxsize = 7.0cm
    \centerline{\epsffile{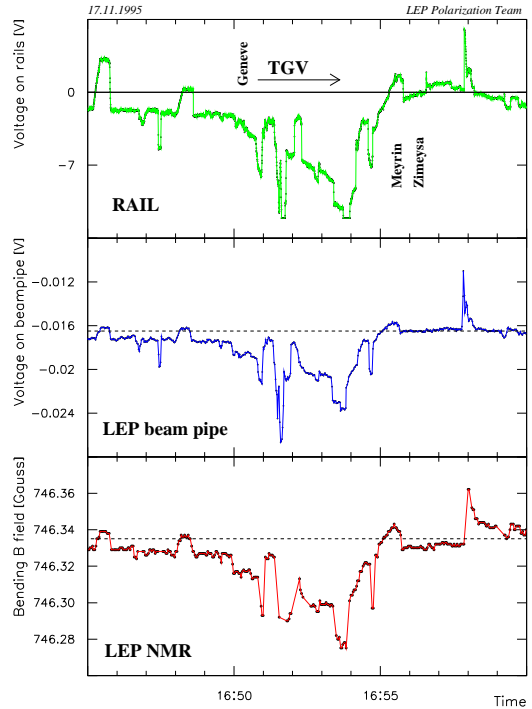}}
\caption[]{From top to bottom: voltage on the rails, voltage on the 
           beampipe and dipole field strength over a period of 15 minutes 
           while the TGV travels closest to the LEP tunnel. }
\label{fig:tgv}
\end{figure}

The calibration of the LEP beam energy is a remarkable feat. The beam 
energy is measured most accurately using the technique of resonant 
depolarization which has an ultimate accuracy of about 200~keV. 
This calibration, however, cannot be performed very often since 
it takes a long time for the transverse beam polarization to build up 
in the accelerator. 
Moreover, it cannot be done during a physics run and has been performed 
with separated beams only. 
The energy of the beam is generally tracked using NMR probes. 
Over the course of the years it was discovered that the circumference  
of the LEP tunnel, and thus the beam energy, was sensitive to the 
water level of Lake Geneva and the phases of the moon. 
The sun and moon tides changed the LEP orbit by up to 1~mm~\cite{lep_ebeam}. 
In 1995 NMR probes were 
installed inside two of the LEP magnets in the tunnel and a new puzzle 
arose. It was observed that there were large fluctuations in the beam 
energy which magically disappeared at midnight only to show up again shortly 
after 4am each day. This effect was eventually traced to an induction 
voltage on the LEP beam pipe caused by vagabond currents on the TGV train 
track rails (see Fig.~\ref{fig:tgv})~\cite{lep_ebeam}. 
All these effects have been taken into account for the results presented
here and propagated back into the previous years with a resulting 
uncertainty on the $Z$ mass and width of 
$\Delta M_Z$ = 1.5~MeV/c$^2$ and 
$\Delta \Gamma_Z$ = 1.7~MeV.

\begin{table*}[t]
\begin{center}
\begin{tabular}{|r||r@{$\pm$}l|r@{$\pm$}l|r@{$\pm$}l|r@{$\pm$}l||r@{$\pm$}l|}
\hline
           &  \mch{ALEPH} & \mch{DELPHI} & \mch{L3} & \mch{OPAL} 
           &  \multicolumn{2}{||c|}{Average Value} \\
\hline
\hline
$\MZ (\GeV)$ & $91.1873$&$0.0030$& $91.1859$&$0.0028$& $91.1883$&$0.0029$
             & $91.1824$&$0.0039$& $91.1863$&$0.0020$  \\ 
$\GZ (\GeV)$ & $2.4950$&$0.0047$ & $2.4896$&$0.0042$ & $2.4996$&$0.0043$
             & $2.4956$&$0.0053$ & $2.4946$&$0.0027$   \\
$\shad (\nb)$& $41.576$&$0.083$  & $41.566$&$0.079$  & $41.411$&$0.074$ 
             & $41.53$&$0.09$    & $41.508$&$0.056$    \\
$R_{e}$      & $20.64$&$0.09$    & $20.93$&$0.14$    & $20.78$&$0.11$   
             & $20.82$&$0.14$    & $20.754$&$0.057$    \\
$R_{\mu}$    & $20.88$&$0.07$    & $20.70$&$0.09$    & $20.84$&$0.10$   
             & $20.79$&$0.07$    & $20.796$&$0.040$    \\
$R_{\tau}$   & $20.78$&$0.08$    & $20.78$&$0.15$    & $20.75$&$0.14$   
             & $20.99$&$0.12$    & $20.814$&$0.055$    \\
$\Afbze$     & $0.0187$&$0.0039$ & $0.0179$&$0.0051$ & $0.0148$&$0.0063$
             & $0.0104$&$0.0052$ & $0.0160$&$0.0024$   \\
$\Afbzm$     & $0.0179$&$0.0025$ & $0.0153$&$0.0026$ & $0.0176$&$0.0035$
             & $0.0146$&$0.0025$ & $0.0162$&$0.0013$   \\
$\Afbzt$     & $0.0196$&$0.0028$ & $0.0223$&$0.0039$ & $0.0233$&$0.0049$ 
             & $0.0178$&$0.0034$ & $0.0201$&$0.0018$   \\
\hline
\hline
\end{tabular}
\caption[]{Line shape and asymmetry parameters from  9-parameter fits
           to the data of the four LEP experiments. The last column gives 
           the LEP averages. }
\label{tab:ninepar}
\end{center}
\end{table*}

The results of the line shape measurements of the four LEP experiments 
are given in the first 6 rows of Table~\ref{tab:ninepar}. The last
column lists the LEP averages\footnote{The determination of 
averages will be discussed in section~\ref{sec:avg}.}.
The accuracy of the measurements is impressive. 
It should be noted that the effects of radiative corrections are applied
within the framework of the $\SM$. For example, initial state 
radiation, which shifts the peak cross section by $\sim$ 89~MeV and 
reduces it by $\sim$~26\%, are taken 
into account through QED radiator functions.


In \pbarp collisions the lineshape of the $Z$ resonance is also probed. 
Because of the large range of
available partonic center of mass energies 
the Drell-Yan process
($\qq \rightarrow (\gamma, Z \rightarrow ) \ \ell^+\ell^- )$ 
can be studied over a large di-lepton invariant mass region.
The invariant mass region well above the $Z$ pole is the region where 
the $\gamma Z$ interference effects are strongest. 
A possible substructure of the partons
would manifest itself most prominently in a modification of the
interference pattern. Substructure of partons is most commonly
parametrized in terms of a contact interaction 
characterized by a phase, $\eta$, leading to constructive ($\eta=-1$) 
or destructive interference ($\eta=+1$) with the $\SM$ Lagrangian, and a 
compositeness scale, $\Lambda_\eta$~\cite{contact}. 
By fitting the di-lepton invariant mass spectrum 
to various assumptions for the compositeness 
scale and the phase of the interference, 
lower limits on the compositeness scale can be set. 

\begin{figure}[t]
    \epsfxsize = 8.0cm
    \centerline{\epsffile{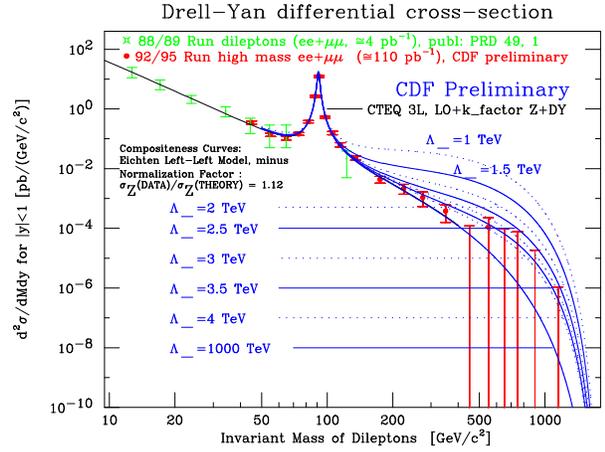}}
\caption{Double differential cross section $d^2\sigma / dM\,dy$
         for CDF electron and muon data combined. The open symbols
         are from the 88/89 data. The solid symbols correspond to the
         full Run I data. The curves are the theoretical predictions
         for different $\Lambda_-$ values. }
\label{fig:cdf_dy}
\end{figure}

The CDF experiment has measured the double differential Drell-Yan
cross section $d^2\sigma / dM\,dy$ for electron and muon pairs
in the mass range $11 < M_{\ell\ell} < 150$ GeV/c$^2$ for the
Run~1a data~\cite{cdf_dy1a}, and
$40 < M_{\ell\ell} < 550$ GeV/c$^2$ for the Run~1b data. 
Figure~\ref{fig:cdf_dy} shows the measured cross section for electrons
and muons combined together with the theoretical predictions. 
The theory curves correspond to a calculation of the
Drell-Yan cross section with in addition a contact interaction of
left-handed quarks and leptons with positive interference for 
different values of the compositeness scale. 
The curve for $\Lambda_-=1000$~TeV indicates the $\SM$ prediction. 
A maximum likelihood fit of the combined electron and muon data 
to the predictions yields 
lower limits in the scale factors of
$\Lambda_+$    $\geq$     2.9 TeV   and
$\Lambda_-$    $\geq$     3.8 TeV.
This implies that up to a distance of $< 10^{-17}$ cm the
interacting particles reveal no substructure.

At $\ee$ colliders particle substructure is also probed using
the angular distribution of the final state leptons in the energy range
$\sqrt{s} = 130 - 140$~GeV with 
similar limits~\cite{delphi_comp}. 

Another very important line shape which yields a mass 
measurement~\cite{lineshape} is the distribution in transverse mass of 
$W\rightarrow \ell\nu$ decays. 
Until very recently the mass of the $W$ boson could only be measured
directly in $\pbarp$ collisions. In a $W$ event originating from a
$\pbarp$ interaction 
in essence only two quantities are measured: the
lepton momentum and the transverse momentum of the recoil system.
The latter consists of the ``hard'' $W$-recoil and the underlying event
contribution, which for $W$-events are inseparable. The transverse momentum
of the neutrino is inferred from these two observables. Since the
longitudinal momentum of the neutrino cannot be determined unambiguously,
the $W$-boson mass is determined using the transverse mass:
\begin{displaymath}
    m_T \,=\,  \sqrt{ 2\, p_T^e \, p_T^\nu \, (1 - \cos\varphi^{e\nu}) } \ ,
\end{displaymath}
where $\varphi^{e\nu}$ is the angle between the electron and neutrino in
the transverse plane.
This distribution exhibits a Jacobian edge, characteristic
of two-body decays, which contains most of the mass information.

As in the measurement of the $Z$ mass, knowledge of the absolute
energy scale is crucial. At LEP the experiments calibrate to the energy 
of the beams, which is known with high precision. The Tevatron
experiments calibrate to known resonances. 
In the CDF $W$-mass analysis~\cite{cdf_mw},
the momentum scale of the central
magnetic tracker is set by scaling the measured $J/\psi$-mass, 
based on an event sample of approximately 60,000 events, to
the world average value using $J/\psi \rightarrow \mu^+\mu^-$ decays.
This procedure establishes the momentum scale at the
$J/\psi$-mass, where the average muon $p_T$ is about 3~GeV/c, and needs to
be extrapolated to the momentum range appropriate for leptons from
$W$-decays. The error due to possible
nonlinearities in the momentum scale is addressed by studying the
measured $J/\psi$-mass as function of $\langle 1/p_T^2 \rangle$,
extrapolated to zero curvature.
Having established the momentum scale, the calorimeter energy scale
is determined from a line shape comparison of the observed $E/p$
distribution with a detailed Monte Carlo prediction. 

In the \D0 $W$-mass analysis~\cite{d0_mw} 
the electromagnetic energy scale is set by calibrating to
the $Z\rightarrow ee$ resonance. The quantity measured is essentially 
the ratio of the measured $W$ and $Z$ mass and the world average
$Z$ mass is used to determine the $W$ boson mass.
By measuring a ratio a number of systematic effects common to both 
measurements cancel. Most notably, 
the ratio is to first order insensitive to the absolute energy scale.
The linearity of the calorimeter is addressed by combining the
measurement of the $Z$ mass with measurements of the decays 
$J/\psi \rightarrow e^+e^-$ and 
$\pi^0 \rightarrow \gamma\gamma \rightarrow e^+e^-e^+e^-$. 


\begin{figure}[t]
\begin{center}
\begin{tabular}{c}
    \epsfxsize = 7.0cm
    \epsffile{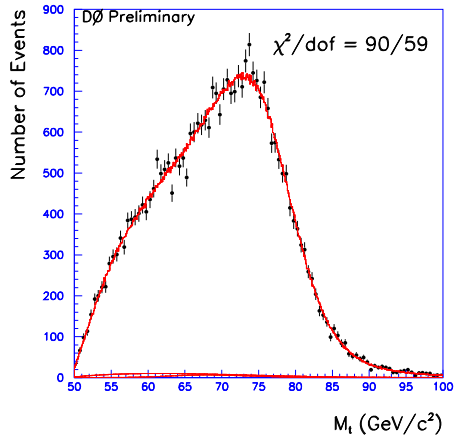}
\end{tabular}
\end{center}
\caption{\D0 transverse mass distribution for $W\rightarrow e\nu$ events
         for the Run~1b data. The points are the data and the line is the 
         best fit to the data. The dashed line indicates the background
         contribution. }
\label{fig:mt_d0}
\end{figure}

Since there is no analytic description of the transverse mass distribution,
the $W$-mass is determined by fitting Monte Carlo generated templates
in transverse mass for different masses of the $W$-boson to the data
distribution. 
Figure~\ref{fig:mt_d0} 
shows the transverse mass distributions for the data
together with the best fit of the Monte Carlo 
for the Run Ib electron data for \D0. 
The $W$ mass is obtained, using central leptons only, 
from a fit in transverse mass over a range 
$ 60 < m_T < 90 $ GeV/c$^2$ for \D0 and 
$ 65 < m_T < 100 $ GeV/c$^2$ for CDF. 
The $W$-mass values obtained are 
$M_W^\mu = 80.310 \pm 0.205 (stat) \pm 0.130 (sys)$~GeV/c$^2$, based on 3268
$W\rightarrow \mu\nu$ events in the mass fitting window, and
$M_W^e   = 80.490 \pm 0.145 (stat) \pm 0.175 (sys)$~GeV/c$^2$, 
based on 5718 events, for CDF using the MRSD$^\prime$- parton
distribution function (pdf). 
\D0 finds
$M_W^e = 80.350 \pm 0.140 \ {\rm (stat.)} \pm 0.165 \ {\rm (syst.)}
                \pm 0.160 \ {\rm (scale)}$~GeV/c$^2$, based on 5982
events in the mass fitting window using the Ia data, and
$M_W^e = 80.380 \pm 0.070 \ {\rm (stat.)} \pm 0.130 \ {\rm (syst.)}
                \pm 0.080 \ {\rm (scale)}$~GeV/c$^2$, based on 27040
events for the Ib data~\cite{d0_mw_1b}. 
Both \D0 measurements are quoted using the MRSA pdf. 
Table~\ref{table:mw_sys} lists the systematic and common errors on the 
measurements.

\begin{table*}[th]
\begin{center}
\begin{tabular}{||l|rrr|rrr||} \hline\hline
    &   \multicolumn{3}{c|}{ CDF }
    &   \multicolumn{3}{c||}{ \D0 }                             \\ \hline
    &   \multicolumn{1}{c}{ e }
    &   \multicolumn{1}{c}{ $\mu$ }
    &   \multicolumn{1}{c|}{ common }
    &   \multicolumn{1}{c}{ Ia }
    &   \multicolumn{1}{c}{ Ib }
    &   \multicolumn{1}{c||}{ common }
\\ \hline
Statistical              & 145   & 205   &  ---  & 140 &  70 & ---    \\
Energy scale             & 120   &  50   &   50  & 160 &  80 &  25    \\
Angle scale              & ---   & ---   &  ---  &  50 &  40 &  40    \\
$E$ or $p$ resolution    & 80    &  60   &  ---  &  70 &  25 &  10    \\
$p_T^W$ and recoil model & 80    &  75   &   65  & 110 &  95 &        \\
pdf's                    & 50    &  50   &   50  &  65 &  65 &  65    \\
QCD/QED corr's           & 30    &  30   &   30  &  20 &  20 &  20    \\
$W$-width                & 20    &  20   &   20  &  20 &  10 &  10    \\
Backgrounds              & 10    &  25   &  ---  &  35 &  15 & ---    \\
Efficiencies             &  0    &  25   &  ---  &  30 &  25 & ---    \\
Fitting procedure        & 10    &  10   &  ---  &   5 &   5 & ---  \\ \hline
Total                    & 230   & 240   &   100 & 270 & 170 &  80    \\
\hline\hline
Combined                 & \multicolumn{3}{c|}  { 180 }
                         & \multicolumn{3}{c||} { 150}                \\
\hline\hline
\end{tabular}
\end{center}
\caption[]{Errors on $M_W$ in \MeV/c$^2$. }
\label{table:mw_sys}
\end{table*}

From the table it can been seen that the error due to the
$p_T^W$ and recoil model and the proton structure are the dominant ones.
The fact that there are spectator interactions, multiple interactions and 
pile-up, with their associated fluctuations and uncertainties, 
is reflected in the recoil 
modeling. It is controlled through the study of $Z$ events and is
expected to scale with the $Z$ statistics. 
The uncertainty due to the parton distribution inside the proton is 
constrained in part by the measurement of the $W$ charge
asymmetry~\cite{cdf_wasym}. 
The CDF experiment uses this measurement
as the sole constraint on the uncertainty due to the $p_T^W$ and \pdfs.
The \D0 experiment addresses the correlation between the 
parton distributions and the spectrum in $p_T^W$ 
by varying both the $p_T^W$ input spectrum and the parton distribution 
functions simultaneously. 
This uncertainty is the dominant theoretical uncertainty which is not 
expected to scale with event statistics. 

Combining~\cite{mw_combined} 
these measurements with previous $W$ mass measurements~\cite{mw_recent},
assuming the only correlated uncertainty between the measurements from 
different experiments is 
due to the parton distribution functions, gives a world average of 
$M_W = 80.356 \pm 0.125$~\GeVcc. 
Since the mass of the $W$-boson is one of the fundamental parameters of the
$\SM$, a precision measurement of the $W$-boson mass 
can be used to look for inconsistencies between the different measurements 
and the theoretical predictions, possibly indicating processes beyond the
$\SM$. The direct $W$ mass measurements will be confronted with 
the prediction from the world's data in section~\ref{sec:all}.

\subsection{Forward-Backward Asymmetry} 

\begin{figure}[h]
    \epsfxsize = 7.0cm
    \centerline{\epsffile{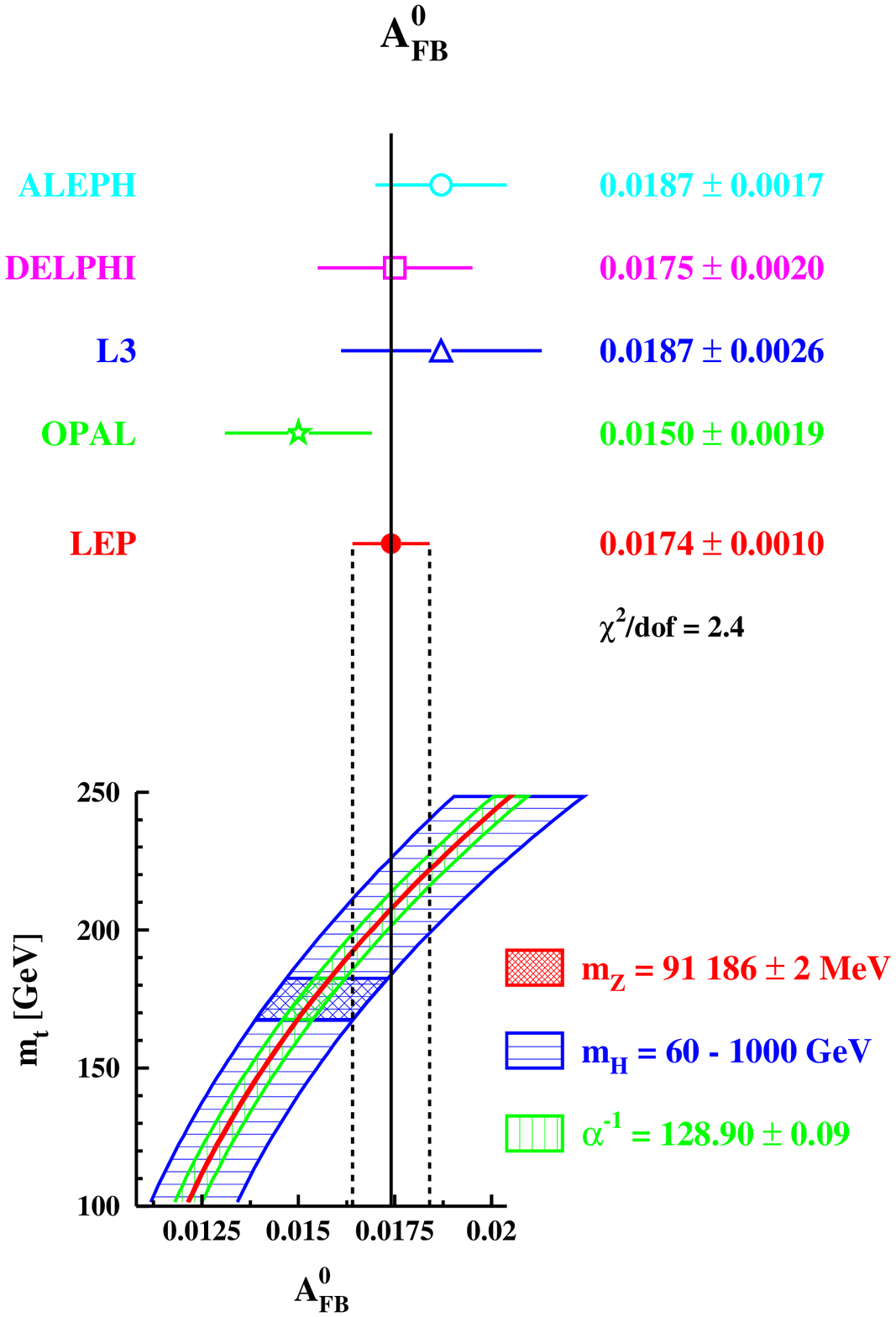}}
\caption[]{$\AFBpole$ measurements of the LEP experiments and their
           average compared to the $\SM$ prediction }
\label{fig:afb}
\end{figure}

The forward-backward asymmetries for leptonic $Z$ decays 
essentially measure the single parameter $\swsqeffl$.
The LEP experiments have measured $\AFB$ both on-pole and off-pole. 
The off-pole measurements are shifted to the pole center of mass energy
using the $\SM$ predicted dependence. This is justified since the slope 
of the asymmetry around $\MZ$ depends only on the axial coupling and the
charge of the initial and final state fermions and is thus independent
of the value of the asymmetry itself.
Figure~\ref{fig:afb} shows the
comparison of the $\AFBpole$ measurements, 
assuming lepton universality, with the $\SM$
prediction. The $\SM$ 
prediction with its uncertainty is given as function of $\Mt$. 
In this figure, and in Fig.~\ref{fig:sweff}, 
three sources of uncertainty on the prediction are 
indicated by bands. Moving outward from the central value they
correspond to the uncertainty on $\MZ$, $\alfmz$ and $\MH$,
respectively. The average value is $\AFBpole = 0.0174 \pm 0.0010$ to be
compared to the $\SM$ prediction of $\AFBpole = 0.0159$\,.

\subsection{Results from Lineshape and Forward-Backward Asymmetry } 
\label{sec:avg} 

\begin{figure}[ht]
    \epsfxsize = 8.0cm
    \centerline{\epsffile{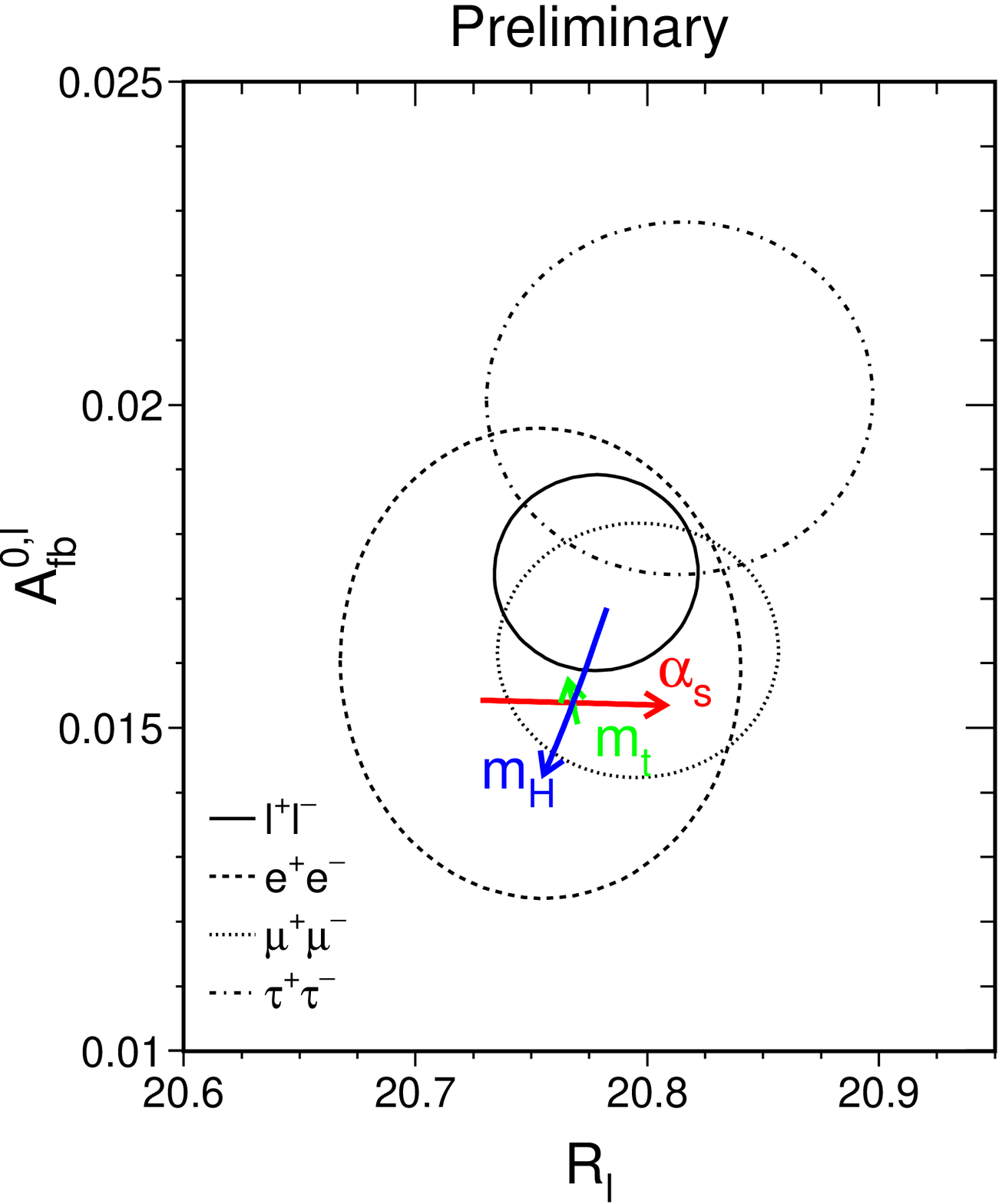}}
\caption[]{Contours of 68\% probability in the $\RZ$-$\AFBpole$ plane. 
           The $\SM$ prediction for $\MZ=91.1863$~\GeV/c$^2$, 
           $\Mt=175$~\GeV/c$^2$, $\MH=300$~\GeV/c$^2$, 
           and $\alphas=0.123$ is also shown. 
           The lines with arrows correspond to the variation of the 
           $\SM$ prediction when $\Mt$, $\MH$ or $\alfmz$ are varied in the 
           intervals $\Mt=175\pm 6$~\GeV/c$^2$,
           $\MH=300^{+700}_{-240}$~\GeV/c$^2$, 
           and $\alfmz=0.123\pm0.006$, respectively. The arrows point in 
           the direction of increasing values of $\Mt$, $\MH$ and
           $\alphas$. }
\label{fig:rl-afb}
\end{figure}

Once the $Z$ lineshape parameters, the forward-backward asymmetries and
the center of mass energies are determined, the results are unfolded for 
initial state radiation and interference effects. That is, 
the $\gamma$-exchange contributions and the $\gamma Z$ interference 
terms are fixed to their $\SM$ values. Each LEP experiment
then performs a fit of the measured quantities in terms of 9 variables, 
$\MZ$, $\GZ$, $\shad$, $\Rl$ and $\Afbzl$. 
This particular choice of variables minimizes the model 
dependence as well as the correlation among them. 
It is the correlation among the parameters that governs which variables
are grouped together in the averaging procedure. 
For example, $\Afbzl$ is strongly dependent on the center of mass energy and
is thus sensitive to initial state radiation and the beam energy. This 
then introduces a correlation with $\MZ$. Therefore, $\Afbzl$ is
included in this particular set of variables for the averaging
procedure. 
The correlations among the different measured quantities is a delicate
matter and a lot of care is given in their determination~\cite{LEP_correl}.
The results among the different experiments are correlated through, for
example, the 
theoretical uncertainty on the luminosity normalization, the uncertainty
on the beam spread and the absolute energy calibration of the beams. 
The results of the 9 parameter fit to the combined LEP data is given in 
the last column of Table~\ref{tab:ninepar}. They are consistent with lepton
universality. The maximum deviation is observed in the $\tau$ sector. 
Assuming lepton universality the
parameter space is reduced to 5 and the results are given in 
Table~\ref{tab:fivepar}. It should be noted that under this assumption 
$\Gll$ in the definition of $\RZ \equiv \Ghad/\Gll$ now refers 
to the partial $Z$ width for the decay into a pair of 
massless leptons. The small mass corrections due to the fermion mass 
are derived within the framework of the $\SM$. 
The results of the lineshape and forward-backward asymmetry measurements
are shown as 68\% probability contours in~Fig.~\ref{fig:rl-afb}. 

\begin{table}[h]
\begin{center}
\begin{tabular}{|c||r@{$\pm$}l|c} \hline
  Parameter     &\mch{ Average Value} \\
\hline
\hline
  $\MZ$ (\GeV/c$^2$)  & $91.1863$ & $0.0020$  \\
  $\GZ (\GeV)$        & $ 2.4946$ & $0.0027$  \\
  $\shad (\nb)$       & $41.508$  & $0.056$   \\
  $\RZ$               & $20.778$  & $0.029$   \\
  $\Afbpole$          & $0.0174$  & $0.0010$  \\
\hline
\end{tabular}
\caption[]{Average line shape and asymmetry parameters from the results
           of the four LEP experiments given in Table~\ref{tab:ninepar}
           assuming lepton universality. } 
\label{tab:fivepar}
\end{center}
\end{table}

The results of the five parameter fit 
can be used to derive the leptonic
and hadronic partial decay widths of the $Z$ boson. 
An important aspect of these measurements is the information relayed 
regarding the invisible $Z$ decay width, given by 
\begin{displaymath}
    \Ginv \, = \,   \GZ \,-\, \Ghad \,-\, (3+\delta_\tau) \,\Gll \ .
\end{displaymath} 
Here $\delta_\tau = -0.0023$ represents a small correction due to the
$\tau$-mass. The measurements give 
$\Ginv /\Gll = 5.952 \pm 0.023$. The $\SM$ predicts 
$\Gnu / \Gll = 1.991 \pm 0.001$, giving for the number of
light neutrino species 
\begin{displaymath}
    N_\nu \,=\, {\Ginv \over \Gll} \, 
                 \left( {\Gll \over \Gnu} \right) 
          \,=\, 2.989 \pm 0.012  \ .
\end{displaymath} 
The advantage here is again the use of ratios. The partial widths 
have a non-negligible top mass dependence due to radiative corrections. 
Since these corrections are mostly universal, 
the dependence is significantly 
reduced in the ratio of partial widths. 
The disadvantage is that the result for the number of light neutrino
species is only valid in the framework of the $\SM$.

\subsection{Polarization} 

The $\SM$ predicts parity violation not only for charged currents 
but for neutral currents as well. For the process $\ee\rightarrow\ff$
it manifests itself through a difference in production cross section 
for fermions with a different polarization. Polarization studies have 
experimentally been approached in two ways. One method, employed by 
the SLC collider, is to polarize the electron beam and 
measure the asymmetry $\ALR$ defined as
\begin{displaymath}
    \ALR \,=\, { \sigma_{\rm L} \,-\, \sigma_{\rm R} \over 
                 \sigma_{tot} } 
\end{displaymath}
where 
$\sigma_{R(L)}$ is the total production cross section for right (left) 
handed polarized electrons. The source of polarized electrons is a 
strained GaAs photocathode, illuminated with circularly polarized light.
Because of the mechanical strain in the solid there is no theoretical 
limitation to the polarization achievable. The SLC polarization group
has steadily improved the polarization over the years reaching an
average polarization during the 1994-1995 run of 
$ {\cal P}_e = (77.34 \pm 0.62)\%$. 
The degree of polarization is measured using a multi-channel 
\v{C}erenkov detector which measures Compton-scattered
electrons from the collision of the longitudinally polarized electron 
beam with a circularly polarized photon beam. The laser polarization 
can be flipped randomly and the asymmetry in cross section is measured. 
Special care has been taken to determine the true luminosity weighted 
polarization for $Z$ production at the interaction point. The Compton
polarimeter measures the polarization of the entire electron bunch. The
machine optics and the inherent beam spread in the bunch, however,
reduce the contribution from off-energy electrons to the $Z$ production
luminosity~\cite{sld_alr94}. 
These effects have all been evaluated and result in a small correction 
of $\sim$0.07\% to the measured polarization. 
The uncertainty on the 
polarization measurement is dominated by the 
uncertainty on the calibration of the \v{C}erenkov detector. 

The measurement of $\ALR$ is relatively straightforward, since it
essentially relies on counting $Z$ events irrespective of their final
state. The measurement is therefore relatively free of systematic
effects. Events from the process $\ee\rightarrow\ee$ are excluded due
to the large zero asymmetry contribution from the $t$-channel diagram. 
At the $Z$ pole, ignoring photonic corrections, $\ALR = \cAe $ 
independent of the final state couplings. 
The SLD collaboration measures~\cite{sld_alr} 
\begin{displaymath}
    \ALRpole \,=\, 0.1542 \pm 0.0037  \ .
\end{displaymath}
where the superscript ``0" indicates that small corrections
have been applied, using the $\SM$ dependencies, to correct for
electroweak interference and pure photon exchange contributions. 
This result yields directly 
\begin{displaymath}
    \swsqeffl \,=\, 0.23061 \pm 0.00047  \ .
\end{displaymath}
It is noteworthy that this single measurement has an accuracy similar 
to the measurement of $\swsqeffl$ from $\AFBpole$ from all LEP experiments 
combined.  
The sensitivities are related as 
$ {\partial\AFB \over \partial\swsqeffl} = {3\over 2} 
   \cAf {1\over {\cal P}_e}
  {\partial\ALR \over \partial\swsqeffl}$. 
Compared to an $\ALR$ measurement using all $Z$ decay channels, 
an approximately 90-fold larger data sample is required to achieve 
a similar accuracy in $\swsqeffl$ from $\AFB$ using leptonic $Z$
decays.

The time-reversal of this process is measured at LEP where the
polarization of the final state particles is measured for unpolarized 
$\ee$-beams: 
\begin{displaymath}
    {\cal P}_f \,=\, { \sigma_R^f \,-\, \sigma_L^f \over 
                       \sigma_{tot} } 
               \,=\, \cAf  \ ,
\end{displaymath}
where $\sigma_R^f$ ($\sigma_L^f$) refers to the production cross section
for right(left)-handed fermions. Similarly to 
$\ALR$ being independent of the final state couplings, the average
polarization of the final state fermions is independent of the initial 
state couplings. 
Because of the
helicity of fermions, ${\cal P}_f$ obviously has an angular dependence 
given by
\begin{eqnarray*} 
    {\cal P}_f (\cos\vartheta)
            &=&  
             - { {\cal A}_{\rm f} (1+\cos^2 \vartheta) \,+\, 
                 2 {\cal A}_e \, \cos\vartheta   
                 \over 
                 1+\cos^2 \vartheta \,+\, 
                 2 {\cal A}_e {\cal A}_{\rm f} \cos\vartheta  }  \ .
\end{eqnarray*} 
This gives rise to a forward-backward polarization asymmetry 
\begin{eqnarray*} 
    A_{\rm {FB}}^{{\cal P}_f}
            \,&=&\, { (\sigma_L^f - \sigma_R^f)_{\rm F}  \,-\, 
                      (\sigma_L^f - \sigma_R^f)_{\rm B}  \over 
                      (\sigma_L^f + \sigma_R^f)_{\rm F}  \,+\, 
                      (\sigma_L^f + \sigma_R^f)_{\rm B} }  
            \\
            \,&=&\, {3\over 4} \, \cAe 
             \,=\,  {3\over 4} \, \langle {\cal P}_Z \rangle  \ ,
\end{eqnarray*} 
which obviously is independent of the final state couplings. 
The forward-backward asymmetry of the fermion polarization is
governed by the average polarization of the $Z$ boson, 
$\langle {\cal P}_Z \rangle$, which depends only on the initial state
couplings. 
The angular distribution of the polarization thus gives independent
measurements of $\cAf$ and $\cAe$, linear in both variables, which 
allows a relative sign determination of $g_V$ and $g_A$. 

The polarization has to date only been measured for $\tau$ leptons 
for which the decay products can be used as spin analyzers assuming the
$V-A$ structure of the weak decay. 
The decays used are 
$\tau\rightarrow \pi\nu_{\tau}$, $\tau\rightarrow \rho\nu_{\tau}$, 
$\tau\rightarrow a_1 \nu_{\tau}$, 
$\tau\rightarrow e\nu_{\tau}\nu_e$ and 
$\tau\rightarrow \mu\nu_{\tau}\nu_\mu$. 
The extraction of the $\tau$ polarization basically employs the 
particle momentum spectrum of the decay particles. The 
$\rho\nu_{\tau}$ and $\pi\nu_{\tau}$ decays contribute most
significantly. 
The LEP measured average values for $\cAt$ and $\cAe$, 
\begin{eqnarray*}
      \cAt &  = & 0.1401 \pm 0.0067 \\
      \cAe &  = & 0.1382 \pm 0.0076 \, ,
\end{eqnarray*}
are compatible with lepton universality. Assuming
$\mathrm{e}-\tau$ universality, the values for $\cAt$ and $\cAe$
can be combined giving $\cAl = 0.1393 \pm 0.0050$\,.

\subsection{Results on Neutral Current Couplings from the Lepton Sector } 
\label{sec:gvga}

\begin{figure}[h]
    \epsfxsize = 7.0cm
    \centerline{\epsffile{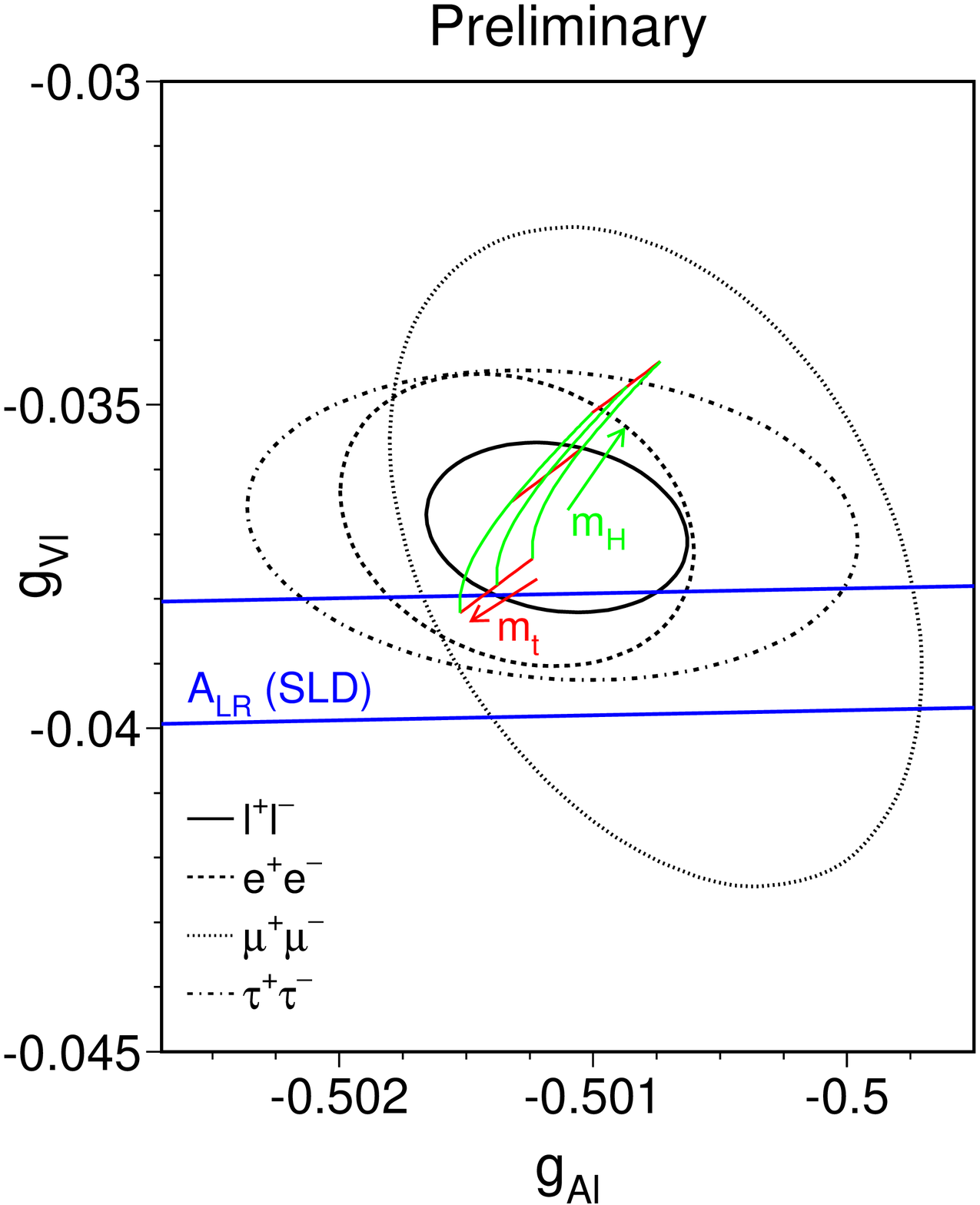}}
\caption[]{Contours of 68\% probability in the $\gvhatl$-$\gahatl$ plane from
           LEP measurements. The grid corresponds
           to the $\SM$ prediction for $\Mt=175\pm 6$~\GeV/c$^2$ 
           and $\MH=300^{+700}_{-240}$~\GeV/c$^2$. }
\label{fig:gagv}
\end{figure}

It is useful at this point to take stock of all the measurements in 
hand. The results on $\Gll$ from the line shape measurements, $\AFBpole$,
$\ptau$, $\AFBtau$ and $\ALR$ are all proportional to 
$\cAl$ or a combination of $\cAl$'s. The results can be combined 
to determine the effective vector and axial-vector coupling constants 
for $e$, $\mu$ and $\tau$ and provides a test of lepton universality. 
Figure~\ref{fig:gagv} summarizes the results as 
contours of 68\% probability in the $\gvhatl$-$\gahatl$ plane from
LEP measurements. The solid contour results from a fit assuming
lepton universality. Also shown is the one standard deviation band
resulting from the $\ALR$ measurement of SLD. The grid corresponds
to the $\SM$ prediction for $\Mt=175\pm 6$~\GeV/c$^2$ and
$\MH=300^{+700}_{-240}$~\GeV/c$^2$. The arrows point, as usual, in the 
direction of increasing value of $\Mt$ and $\MH$.
The average central values are given in Table~\ref{tab:gvga}.
The neutrino coupling to the $Z$ is derived from the
measured value of its invisible width, $\Ginv$, attributing it
exclusively to the decay into three identical neutrino generations
($\Ginv=3\Gnu$) and assuming $\gahatn =\gvhatn =\ghatn$.

\begin{table}[ht]
\begin{center}
\begin{tabular}{|l|c|c|}
\hline
 & LEP & LEP+SLD\\
\hline
\hline
$\gvhatl$    & $-0.03688  \pm 0.00085 $ & $-0.03776  \pm 0.00062 $ \\
$\gahatl$    & $-0.50115  \pm 0.00034$  & $-0.50108  \pm 0.00034 $ \\
$\ghatn$     & $+0.5009   \pm 0.0010 $  & $+0.5009   \pm 0.0010  $ \\
\hline
\end{tabular}
\end{center}
\caption[]{Results for the effective vector and axial-vector
           couplings from the combined LEP data assuming lepton 
           universality. For the right column the SLD measurement 
           of $\ALR$ has been included.}
\label{tab:gvga}
\end{table}

\section{Heavy Flavor Sector } 
Of particular interest in the heavy flavor sector are the 
ratios of the $b$ and $c$ quark partial widths of the $Z$ to the 
total hadronic partial width, $\Rb \equiv \Gbb / \Ghad$ and 
$\Rc \equiv \Gcc / \Ghad$, respectively. 
Because the $b$ quark is in the same isospin doublet as the $t$ quark,
the $Z\rightarrow \bb$ partial width receives vertex corrections which
are unique to this particular decay mode and is thus very sensitive to
physics beyond the $\SM$. 
For a long time both $\Rc$ and $\Rb$ deviated substantially from 
the $\SM$ prediction. At the 1995 summer conferences 
the values reported were 
$\Rc$ = 0.1543 (74) and 
$\Rb$ = 0.2219 (17), compared to their $\SM$ values of 
0.1724 and 0.2156, respectively~\cite{ichep95}. 
Taken at face value, assuming Gaussian errors, the $\Rb$ measurement 
ruled out the $\SM$ at more than 99.9\% CL, and excited tremendous 
interest among theorists proposing all kinds of 
extensions to the $\SM$~\cite{leptophobia}. 
Because the new results for $\Rb$ and $\Rc$ have 
changed significantly, the focus of this section will be on the 
new measurements of these two quantities. 

\subsection{$\Rc$ } 
The $\Rc$ and $\Rb$ analyses rely on the identification of 
events as originating from the decay of a $c$ or $b$ quark, called 
\lq\lq tagging\rq\rq, with a minimal background and small hemisphere 
correlations. The oldest method to
tag events employs the lepton $p_T$ spectrum of the semi-leptonic
decays of the heavy quarks. 
Two new methods to tag $c$ quark events for the $\Rc$ measurement have 
been developed based on \lq\lq charm counting\rq\rq\ and tagging charm 
events using a \lq\lq slow\rq\rq\ pion. 

\begin{figure}[h]
    \epsfxsize = 8.0cm
    \centerline{\epsffile[25 200 525 650]{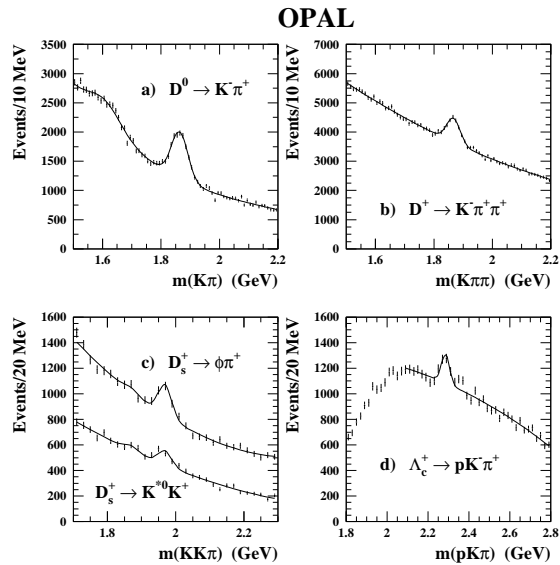}}
\caption{OPAL mass distributions of the five reconstructed D~meson decays. }
\label{fig:c_count}
\end{figure}

The charm counting method is based on the observation that all charm
quarks end up in the weakly decaying charmed hadrons\footnote{Charge
conjugation is implied throughout in this section.}
$\Dzero$, $\Dplus$, $\Ds$ and $\Lc$: 
\begin{eqnarray*}
    {\rm P}(c \rightarrow \Dzero)  \,+\, 
    {\rm P}(c \rightarrow \Dplus)  \,+\, 
    {\rm P}(c \rightarrow \Ds)     \,+\,    \\
    {\rm P}(c \rightarrow \Lc)    (1 \,+\, S_{\rm baryon}) \,=\, 1
\end{eqnarray*}
Here ${\rm P}( c \rightarrow {\rm X}_c)$ is the probability that a
primary $c$ quark results in the production of charmed hadron 
${\rm X}_c$. 
$S_{\rm baryon}$ is a correction factor of 0.15 for the formation of 
strange-charmed baryons, like $\Xi^+_c$. The charmed hadrons are
reconstructed in the decay modes 
\begin{center}
\begin{tabular}{lcl} 
    $\Dzero$ & $\rightarrow$  &  ${\rm K}^-\pi^+$                   \\
    $\Dplus$ & $\rightarrow$  &  ${\rm K}^-\pi^+\pi^+$              \\
    $\Ds$    & $\rightarrow$  &  $\phi\pi^+$                        \\
    $\Ds$    & $\rightarrow$  &  ${{\overline {\rm K}}^*}^0 {\rm K}^+$ \\
    $\Lc$    & $\rightarrow$  &  ${\rm pK}^-\pi^+$                 
\end{tabular}
\end{center}
Figure~\ref{fig:c_count} shows the mass distributions from the OPAL
experiment for the five decay modes~\cite{opal_c}. 
These event samples are certainly 
not free of charmed hadrons from $b$ decays. 
The relatively large $b$
hadron lifetimes and hard $b$ fragmentation result in significantly
longer apparent decay lengths and softer energy spectra for these
charmed hadrons compared to those from primary charm production. 
This provides handles to separate the contributions from $b$ hadron
decays and from prompt production. 
The overall contribution from $b$ decays in the event sample, however, 
still exceeds that from primary $c$ decays and the results are sensitive
to uncertainties in $b$ fragmentation and $b$ hadron lifetimes. 
These are the dominant systematic uncertainties and have been addressed 
by Monte Carlo. 
The reconstruction efficiencies for each of the separate decays have 
also been determined by Monte Carlo. 
They are slightly lower for primary charmed hadrons than 
for charmed hadrons coming from $b$ decays. 
An important additional source of background is the production of 
charmed hadrons through gluon splitting, $g\rightarrow c{\overline c}$. 
Although the event selection is geared towards selecting energetic
D~mesons, about half of all the D~mesons from gluon splitting remain in
the event sample. The mean multiplicity of $c{\overline c}$ production
from gluon splitting in hadronic $Z$ decays was measured from the
production of $\Dstar$ mesons to be 
${\overline n}_{g\rightarrow c{\overline c}} = (4.4 \pm 2.1)\%$
\cite{opal_dstar}. Recent measurements based on leptonic events yield 
${\overline n}_{g\rightarrow c{\overline c}} = (2.38 \pm 0.48)\%$, 
thus raising $\Rc$ since less charm background from gluon splitting is
subtracted.

Knowing the efficiencies and the background contributions the data
allows for a direct measurement of $\RcPcX$. 
With the constraint that the probabilities for the weakly
decaying charmed hadrons add up to one, the sum of these 
measurements, corrected for the decay branching ratios as listed by the
particle data group, yields $\Rc$. 
It is important to note that no assumptions need to be made on the 
production rates of the individual charmed mesons. 
The analyses do depend, however, on the measured branching ratios, 
which are used as an external input. 
Because of a new measurement by ARGUS~\cite{argus}, the average 
branching ratio 
$BR(\Dzero\rightarrow {\rm K}^-\pi^+)$ has changed significantly from 
(4.01$\pm$0.14)\% to (3.83$\pm$0.12)\%~\cite{pdg}, also resulting in an 
increase in $\Rc$. 

\begin{figure}[h]
    \epsfxsize = 7.0cm
    \centerline{\epsffile{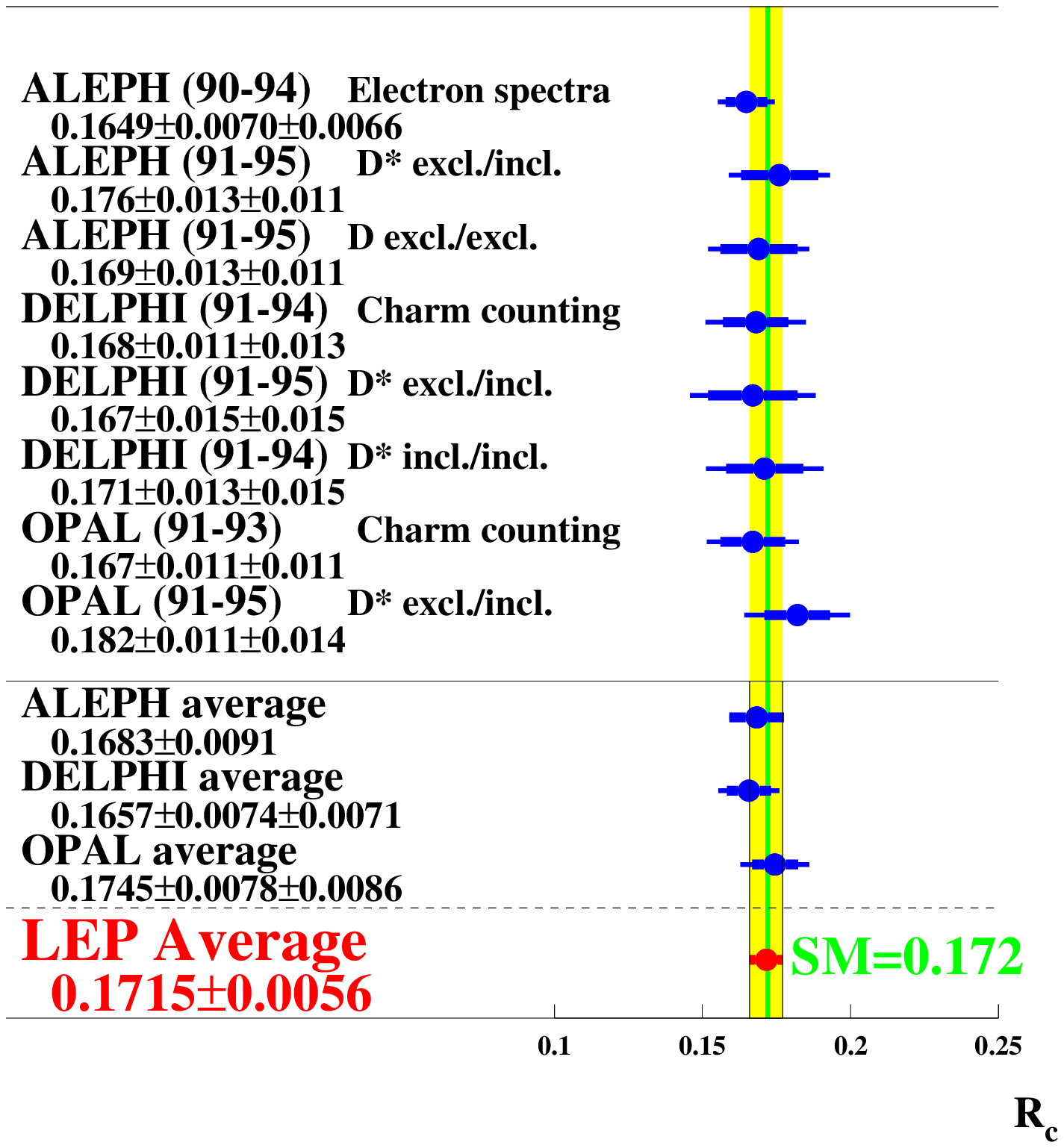}}
\caption{Summary of $\Rc$ measurements. }
\label{fig:rc}
\end{figure}

Alternative methods to measure $\Rc$ use the decay 
$\Dstar \rightarrow \Dzero\pi^+ \rightarrow ({\rm K}^-\pi^+)\pi^+$. 
Because of the very low $Q$ value of the decay, the pion from the $\Dstar$
decay has a very low $p_T$ with respect to the $\Dstar$ line of flight 
and can be used to tag the event. 
The slow pion tag analyses generally proceed by 
first measuring the production rate of single tagged, exclusive $\Dstar$
decays, $N_d$, given by 
\begin{eqnarray*} 
    {N_d \over N_{had} } &\sim &  \Rc \cdot
                                   P(c\rightarrow \Dstarp) 
                                   BR(\Dstarp) \  \epsilon_{\Dstarp}     
\end{eqnarray*} 
where $N_{had}$ is the number of hadronic $Z$ decays, and 
$\epsilon_{\Dstarp}$ the $\Dstarp$ reconstruction efficiency. 
In a second step an inclusive \lq\lq slow\rq\rq\ pion tag is applied 
to the opposite hemisphere giving for the number of double tagged events, 
$N_{dd}$, 
\begin{eqnarray*} 
    {N_{dd} \over N_{had} }  
            &\sim & \Rc \cdot \left[
                     P(c\rightarrow \Dstarp) BR(\Dstarp) \right]^2 
                    \,  \epsilon_{\Dstarp}  
                    \,  \epsilon_{s} 
\end{eqnarray*} 
with $\epsilon_{s}$ the slow pion tag efficiency.
Each experiment has its own variant of
this procedure. DELPHI, for example, 
uses a fully inclusive 
tag, with high efficiency and large backgrounds~\cite{rc_delphi}. 
ALEPH~\cite{rc_aleph} and OPAL~\cite{rc_opal} use an 
inclusive--exclusive tag using 
$\Dstar$ mesons, although ALEPH has also tried a fully exclusive tag 
of D-meson decays with reduced statistics but much higher purity. 
An important bonus of these analyses is that $\PcDst$ is measured 
directly and does not 
need to be taken from low-energy data as external input. 
A summary of all $\Rc$ results from the different measurement
techniques is shown in Fig.~\ref{fig:rc}. 
                                                                        
To summarize, the main reasons for the increase in $\Rc$ are: 
{\it i)}    more data analysed, 
{\it ii)}   decrease in the gluon splitting probability 
            $g\rightarrow c{\overline c}$, 
{\it iii)}  decrease in the branching ratio 
            $BR(\Dzero\rightarrow {\rm K}^-\pi^+)$ and 
{\it iv)}   new Aleph measurement. 
The new value of $\Rc$ is in excellent agreement with the $\SM$
prediction. The change is dominated by the updated OPAL
measurement and the new ALEPH result.

%
%

\subsection{$\Rb$ } 
The measurements of $\Rb$ also employ the single-tag and double-tag
technique. As noted in the measurement of $\Rc$, in the single tag 
method the number of tagged events is counted. This
number is corrected for backgrounds from other flavors and
for the tagging efficiency to calculate the true fraction of hadronic
$Z$ decays of that flavor. 
For the double-tag measurement, the event is divided into two
hemispheres and both hemispheres are tagged. 
Writing the number of tagged
single hemispheres as $N_t$, the number of events with both
hemispheres tagged as $N_{tt}$, then for a total of $N_{\rm{had}}$
hadronic $Z$ decays the measurement of $\Rb$ follows from 
\begin{eqnarray}
   \frac{N_t}{2N_{\rm{had}}}   &=& \effb \Rb  +
                                   \effc \Rc  +
                                   \effuds ( 1 - \Rb - \Rc ) ,
\nonumber \\
   \frac{N_{tt}}{N_{\rm{had}}} &=& (1 + \rho_b) 
                                   \effb^2 \Rb  +
                                   \effc^2 \Rc  + 
\nonumber \\
                               & & \effuds^2 ( 1 - \Rb - \Rc ) ,
\label{eq:btag} 
\end{eqnarray}
where $\varepsilon_{b}$, $\varepsilon_{c}$ and 
$\varepsilon_{uds}$ are the tagging efficiencies per
hemisphere for $b$, $c$ and light-quark events, and $\rho_b$
accounts for the fact that the tagging efficiencies between the
hemispheres may be correlated.  
By measuring both the single and double tag rate, the $b$ tagging
efficiency can be determined directly from the data, reducing the 
systematic uncertainties in the measurement. 

The most precise determinations of $\Rb$ use the lifetime tag of 
the $b$-quark. Events are tagged by reconstructing either a secondary 
vertex (SV) or an impact parameter. Events originating from $b$ decays
will have large positive values for these quantities. The negative
tails in these distributions are used to measure the resolutions and 
control systematic effects. 
The measurements of $\Rb$ were, and still are, systematics dominated. 
Two of the dominant sources of systematics are the charm background 
and the hemisphere correlations. The experimental effort therefore has
gone into reducing both $\effc$ and $\rho_b$ in equation~(\ref{eq:btag}). 
It should be pointed out here that the correlations are analysis
dependent and are very different for an impact parameter analysis
compared to a measurement using the SV technique. 

\begin{figure}[h]
    \epsfxsize = 7.0cm 
    \centerline{\epsffile{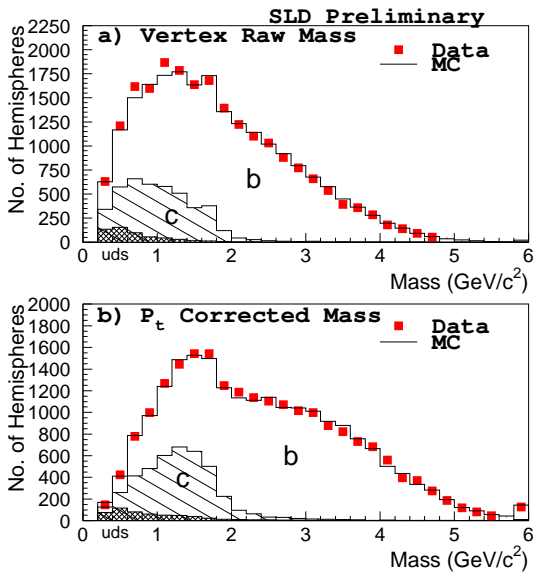}} 
\caption{a) Secondary vertex mass distribution as measured by SLD 
            (points) together with $u, d, s$ (cross hatched) and 
            charm (hatched) background contributions. 
         b) Mass distribution corrected for missing transverse 
            energy. } 
\label{fig:sld_mass}
\end{figure}

\begin{figure}[h]
    \epsfxsize = \linewidth 
    \centerline{\epsffile{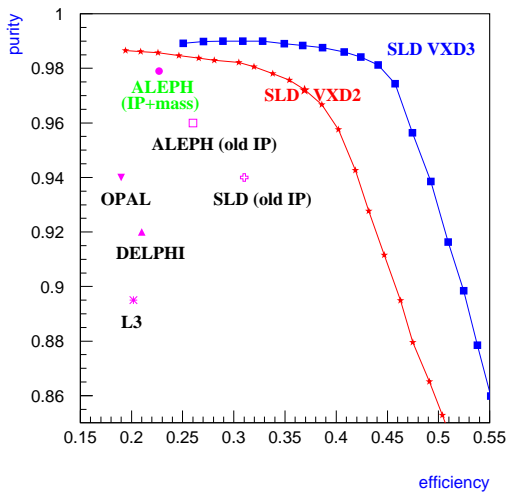}}
\caption{$b$ tag performance per hemisphere for the different 
         experiments. }
\label{fig:b_perf}
\end{figure}

As for the charm sector, there have been two significant developments. 
First of all,
charm decays are currently much better understood by the LEP and SLD 
experiments. The production rates of the different charmed mesons and
the branching ratios of cascade decays, for example, are now measured by
the experiments themselves. Secondly, a new lifetime and mass tag,
first presented by SLD~\cite{rb_sld} with a similar method 
developed independently by ALEPH, 
has allowed for 
a substantial reduction of the charm background in the data sample. 
This tag proceeds by first computing the confidence level that all
tracks in a hemisphere come from the primary vertex (PV). 
Tracks least consistent with the PV are then combined and their 
invariant mass calculated. Figure~\ref{fig:sld_mass}a show this mass
spectrum as measured by SLD. A cut is 
placed at approximately the charm threshold to obtain the $b$ rich
sample. 
Since the interaction point is very well known at SLC, the SLD
experiment can take this method one step further and correct for the
undetected neutrals in the $b$ decay. A correction is applied to correct
for the missing energy transverse to the direction of flight of the 
$b$ hadron, as given by the PV and SV 
(Fig.~\ref{fig:sld_mass}b). A cut on this \lq\lq $p_T$ corrected 
vertex mass\rq\rq\ is applied to further enrich the sample. 
Due to the larger spread in beam size at LEP, this correction cannot be 
applied by the LEP experiments. 
Note that the presence of charm background in the sample gives rise to an
explicit correlation between $\Rb$ and $\Rc$. 
Figure~\ref{fig:b_perf} summarizes the $b$ tagging performance of the
different experiments. They all 
reach an impressive purity with good detection efficiencies. 

%
%

\begin{figure}[h]
    \epsfxsize = 8.0cm
    \centerline{\epsffile{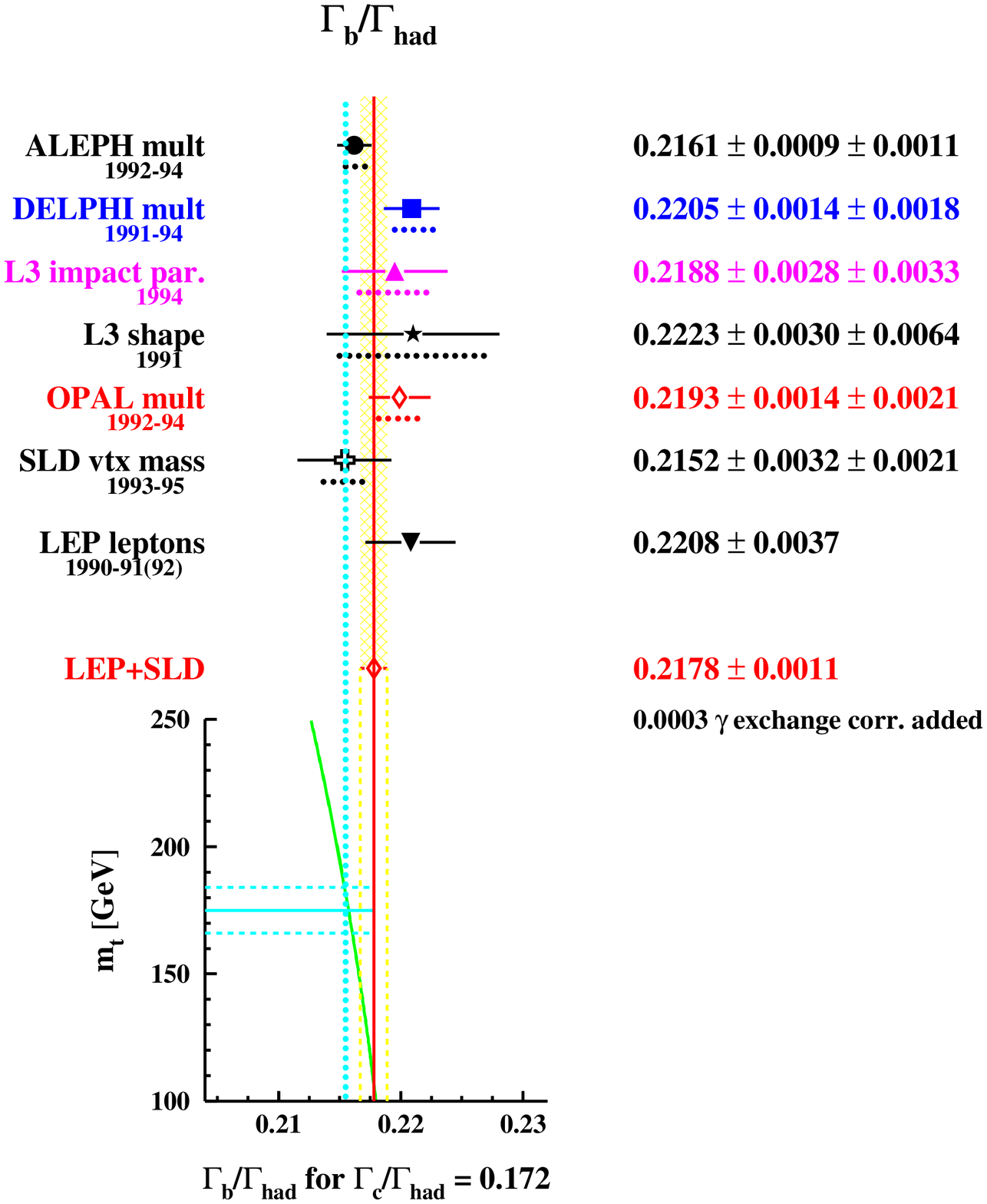}}
    \caption{Summary of $\Rb$ measurements. }
\label{fig:rb}
\end{figure}

There has also been considerable progress in the understanding of
hemisphere correlations. These correlations arise mainly from the 
primary vertex, and from detector and QCD effects. 
If, for example, 
one $b$ hadron has a very long lifetime, 
the efficiency for tagging the other $b$ will be decreased due 
to the degraded PV resolution. 
As most $b$ hadrons are roughly back to back, detector correlations 
are introduced if a region of poorer instrumentation is hit. 
The ALEPH experiment has switched to a method where a PV 
is calculated for each hemisphere, thereby eliminating one of the
dominant contributions to $\rho_b$. 
An alternative method, pioneered by DELPHI, 
employs multiple mutually exclusive tags using the lifetime-mass
information as well as event shape variables. 
The determination of the correlations and their effect on the measurement
is complicated. They are evaluated using both data and Monte Carlo. 
It are these correlations as well as the residual background of other 
flavors which are still the main sources of systematic uncertainty. 
A new ALEPH measurement, using the full 91--95 statistics, has 
currently the smallest error of all individual measurements. It is 
based on multiple mutually exclusive tags using event shape and 
lifetime-mass information and gives 
$\Rb = 0.2158 \pm 0.0009 \pm 0011$, using the $\SM$ value for
$\Rc$~\cite{rb_aleph}. 
In addition to this new measurement, DELPHI has updated its measurements
by inclusion of the 1994 data~\cite{rc_delphi} and L3 has for the first time
presented a lifetime tag measurement~\cite{rb_l3}. 
All results are summarized in Fig.~\ref{fig:rb}. 
The combined LEP/SLD average is $\Rb = 0.2178 \pm 0011$ ($\Rc = 0.172$) 
to be compared to the $\SM$ prediction of $\Rb = 0.2158$\,. 

In summary, the main reasons for the decrease in $\Rb$ are: 
{\it i)}    inclusion of much more data, 
{\it ii)}   better understanding of the charm sector, 
{\it iii)}  reduction of the charm background and 
{\it iv)}   a better understanding of the hemisphere correlations. 
All effects have the
tendency to lower $\Rb$, though the change is dominated by inclusion of 
new data. The change in external input parameters results in a change 
in $\Rb$ of only 0.0003\,. 

\subsection{Other Heavy Flavor Results. } 
The measurements in the heavy flavor sector encompass many other
results. The forward-backward asymmetries for $b$ and $c$ quarks 
measured on- and off-pole, the semi-leptonic branching
ratios $\BrblX$, $\BrclX$, the average $\bb$ mixing parameter 
$\chi$, the various production probabilities for D-mesons
and the quark coupling parameters $\cAb$ and $\cAc$ are all
measured. 
The latter two are measured directly by SLD from the polarized 
forward-backward asymmetry: 
\begin{eqnarray*} 
    A_{\rm {FB}}^{\rm {pol}(f)}
              &=&   { (\sigma_L^f - \sigma_R^f)_{\rm F}  \,-\, 
                      (\sigma_L^f - \sigma_R^f)_{\rm B}  \over 
                      (\sigma_L^f + \sigma_R^f)_{\rm F}  \,+\, 
                      (\sigma_L^f + \sigma_R^f)_{\rm B} }  
            \,=\,   {3\over 4} \, \cAf \ .
\end{eqnarray*} 
Three different 
techniques are used to measure $\cAb$ based on the determination of the 
jet-charge, tagging events through their lepton $p_T$ spectrum and 
tagging with K$^\pm$ mesons~\cite{ab_sld}. 
These analyses have similar sources of systematic error compared to 
the LEP asymmetry measurements.
The SLD measurements yield 
\begin{eqnarray*}
  \cAb    &=& 0.863  \pm  0.049      \\
  \cAc    &=& 0.625  \pm  0.084  \ .  
\end{eqnarray*}
All in all, 17 variables are measured in the heavy flavor sector. 
This set is reduced by four by shifting the off-pole forward-backward
asymmetries to the pole center of mass energy. 
In a fashion similar to the results from the lepton sector, the averages
of all measurements have been determined taking into account their 
correlations~\cite{lep_hf}.

\section{Combining All Results } 
\label{sec:all}

It is widely anticipated that the $\SM$ is just an approximate theory
and should eventually be replaced by a more complete and fundamental
description of the underlying forces in nature. 
The individual measurements probe different aspects of the $\SM$ and 
all measurements combined provide a powerful constraint. 
To test how well the $\SM$ fares one first determines how well the
individual measurements can be accommodated within its framework. 
If they are all consistent, the measurements can be combined to provide 
constraints on those parameters that enter via radiative corrections. 
These constraints can then be compared with direct measurements, if they
exist. This can be an iterative process in which more and
more measurements are included in the full set of electroweak
measurements in each subsequent step. 
In the following subsections the results of taking these successive
steps will be described.

\subsection{The Effective Electroweak Mixing Angle $\swsqeffl$}

\begin{figure}[h]
    \epsfxsize = 8.0cm 
    \centerline{\epsffile{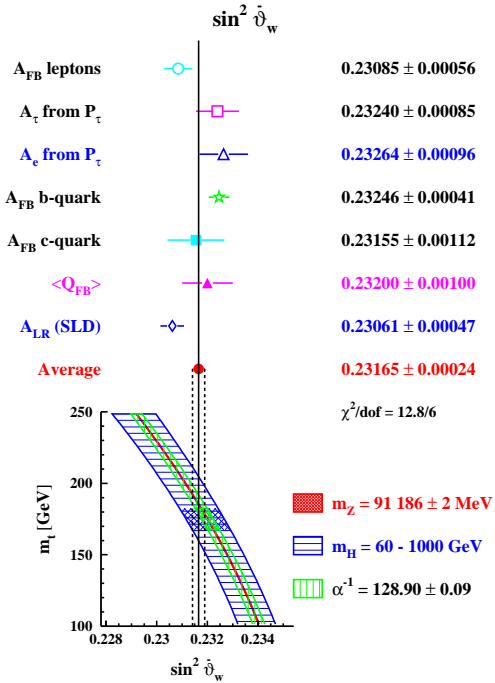}}
    \caption{Summary of $\swsqeffl$ measurements from the forward-backward 
             asymmetries of leptons, $\tau$ polarization, inclusive quarks, 
             heavy quark asymmetry and the SLD polarization asymmetry. } 
    \label{fig:sweff}
\end{figure}

In section~\ref{sec:gvga} the results on $\Gll$, $\AFBpole$,
$\ptau$, $\AFBtau$ and $\ALR$ were combined 
to determine the effective vector and axial-vector coupling constants. 
All asymmetry measurements can be combined into a single
observable, the effective electroweak mixing angle. 
For a combined average of $\swsqeffl$ from $\AFBpole$, $\cAt$, $\cAe$ 
and $\ALR$ only the assumption of lepton universality, already inherent
in the definition of $\swsqeffl$, is needed. Also the 
quark forward-backward asymmetries, $\Afbzb$ and $\Afbzc$, and the 
forward-backward asymmetry in mean jet charge, $\avQfb$, are included in 
this average, as these asymmetries have a reduced sensitivity to 
corrections particular to the hadronic vertex. 
Figure~\ref{fig:sweff} shows the comparison of the individual
measurements with the $\SM$ prediction.
It is seen that there is good agreement between the average of 
$\swsqeffl = 0.23165 \pm 0.00024$, a 0.1\% measurement, 
with the $\SM$ prediction of 
$\swsqeffl = 0.23167$. 
It should be noted that the SLD value for $\swsqeffl$ from $\ALR$ is 
2.2 standard deviations low compared to the world average. 
Most of that discrepancy comes from the early SLD data.

\begin{table*}[th]
\begin{center}
{\footnotesize 
\begin{tabular}{|ll|r|r|r|}
\hline
 &&&&  \\[-3mm]
 && \mcc{Measurement with}  & \mcc{Standard} & \mcc{Pull} \\
 && \mcc{Total Error}       & \mcc{Model}    &  \\
 &&&& \\[-3mm]
\hline
a) & \underline{LEP}     &&& \\
   &                     &&& \\[-3mm]
   & line-shape and      &&& \\
   & lepton asymmetries: &&& \\
&$\MZ$ [\GeV/c$^2$] & $91.1863\pm0.0020\pz$ & 91.1861$\pz$ & $ 0.1$ \\
&$\GZ$ [\GeV{}]     & $2.4946 \pm0.0027\pz$ &  2.4960$\pz$ & $-0.5$ \\
&$\shad$ [nb]       & $41.508 \pm0.056\pzz$ & 41.465$\pzz$ & $ 0.8$ \\
&$\RZ$              & $20.778 \pm0.029\pzz$ & 20.757$\pzz$ & $ 0.7$ \\
&$\Afbpole$         & $0.0174 \pm0.0010\pz$ &  0.0159$\pz$ & $ 1.4$ \\
&+ correlation matrix                            &&& \\
&                                                &&& \\[-3mm]
&$\tau$ polarization:                            &&& \\
&$\cAt$         & $0.1401\pm 0.0067\pz$ & 0.1458$\pz$ & $-0.9$ \\
&$\cAe$         & $0.1382\pm 0.0076\pz$ & 0.1458$\pz$ & $-1.0$ \\
&                       &&& \\[-3mm]
&b and c quark results: &&& \\
&$\Rb$           & $0.2179\pm0.0012\pz$ & 0.2158$\pz$ & $ 1.8$ \\
&$\Rc$           & $0.1715\pm0.0056\pz$ & 0.1723$\pz$ & $-0.1$ \\
&$\Afbzb$        & $0.0979\pm0.0023\pz$ & 0.1022$\pz$ & $-1.8$ \\
&$\Afbzc$        & $0.0733\pm0.0049\pz$ & 0.0730$\pz$ & $ 0.1$ \\
&+ correlation matrix                         &&& \\
&                                             &&& \\[-3mm]
&$\qq$ charge asymmetry:                      &&& \\
&$\swsqeffl$
  ($\avQfb$)   & $0.2320\pm 0.0010\pz$ & 0.23167     & $ 0.3$ \\
&                   &&& \\[-3mm]
\hline
b) & \underline{SLD} &&& \\
   &                 &&& \\[-3mm]
&$\swsqeffl$ ($\ALR$)
             & $0.23061 \pm 0.00047  $ &0.23167      & $-2.2$ \\
&$\Rb$ 
             & $0.2149  \pm 0.0038\pz$ &0.2158$\pz$  & $-0.2$ \\
&$\cAb$ 
             & $0.863   \pm 0.049\pzz$ &0.935$\pzz$  & $-1.4$  \\
&$\cAc$ 
             & $0.625   \pm 0.084\pzz$ &0.667$\pzz$  & $-0.5$ \\
&                                &&& \\[-3mm]
\hline
c) & \underline{$\pbarp$ and $\nu$N} &&& \\
   &                              &&&  \\[-3mm]
&$\MW$ [\GeV/c$^2$] ($\pbarp$)
                 & $80.356\pm 0.125\pzz   $
                                &  80.353$\pzz$& $ 0.3$ \\
&$\swsq$     ($\nu$N)
                 & $0.2244  \pm 0.0042\pz$
                                & 0.2235$\pz$  & $ 0.2$ \\
&$\Mt$ [\GeV/c$^2$] ($\pbarp$)
                 & $175\pm 6\pzz\pzz$
                                & 172$\pzz$    & $ 0.5$ \\
&                                 &&&  \\
\hline
\end{tabular}
}
\end{center}
\caption[z]{Summary of measurements included in the combined analysis of 
            $\SM$ parameters. Section~a) summarizes LEP averages, 
            Section~b) SLD results and Section~c) electroweak
            measurements from $\pbarp$ colliders and $\nu$N scattering. 
            The $\SM$ results in column~3 and the difference between 
            measurement and fit in units of the total measurement error 
            in column~4 are derived from the $\SM$ fit including 
            all data with the Higgs mass treated as a free parameter. }
\label{tab:summary}
\end{table*}

\subsection{The Coupling Parameters $\cAf$ }

The (polarized) forward-backward asymmetry measurements all measure 
either the product of coupling parameters $\cAf$ of different fermion 
species or the single coupling directly. Also the measurement of the 
$\tau$-polarization determines $\cAt$ and $\cAe$, separately. 
Assuming lepton universality, $\cAl$ as determined from 
$\AFBpole$, $\ptau(\cos\vartheta)$ and $\ALR$ is 
\begin{eqnarray*}
    \cAl &=& 0.1466 \pm 0.0033    \qquad {\rm (LEP)}      \\    
    \cAl &=& 0.1500 \pm 0.0025    \qquad {\rm (LEP+ SLD)}    
\end{eqnarray*}
Note that $\cAl$ is pushed up by one standard deviation by
inclusion of the SLD $\ALR$ measurement. 
Using these values for $\cAf$ 
the couplings for the heavy flavors can be
determined from $\Afbzb$ and $\Afbzc$ and the heavy flavor 
left-right asymmetries from SLD. 
Taking the LEP average for $\cAf$ gives 
\begin{eqnarray*}
    \cAb &=& 0.890 \pm 0.029     \\    
    \cAc &=& 0.667 \pm 0.047     
\end{eqnarray*}
whereas using the combined LEP/SLD result for $\cAf$ gives 
\begin{eqnarray*}
    \cAb &=& 0.867 \pm 0.022     \\    
    \cAc &=& 0.649 \pm 0.040     \ , 
\end{eqnarray*}
moving $\cAb$ down by about one standard deviation. 
$\cAc$ agrees very well with the $\SM$ prediction of 0.667. The world
average value for $\cAb$, however, deviates by 3.1 standard deviations
from the $\SM$ prediction of 0.935. 
This deviation is not without controversy. 
It should be kept in mind that the value for $\cAb$ as obtained above is
not an independent measurement since it uses the value for $\cAe$.
Fluctuations in the measurement of $\cAe$, a measurement which is 
unrelated to the b-coupling per se, increase the deviation of $\cAb$ 
with the $\SM$ prediction. 
There is only one direct measurement of $\cAb$, namely from the
left-right forward-backward asymmetry measurement by SLD, 
$\cAb = 0.863 \pm 0.049$, which is 1.4 standard deviations low compared
to the $\SM$ value. 
If one wishes to combine different measurements a value less prone to 
fluctuations in other measurements may be obtained by using the $\SM$ 
prediction for~$\cAe$.

\subsection{Constraints on the Standard Model }

\begin{figure}[h]
        \epsfxsize = 8.0cm 
        \centerline{\epsffile{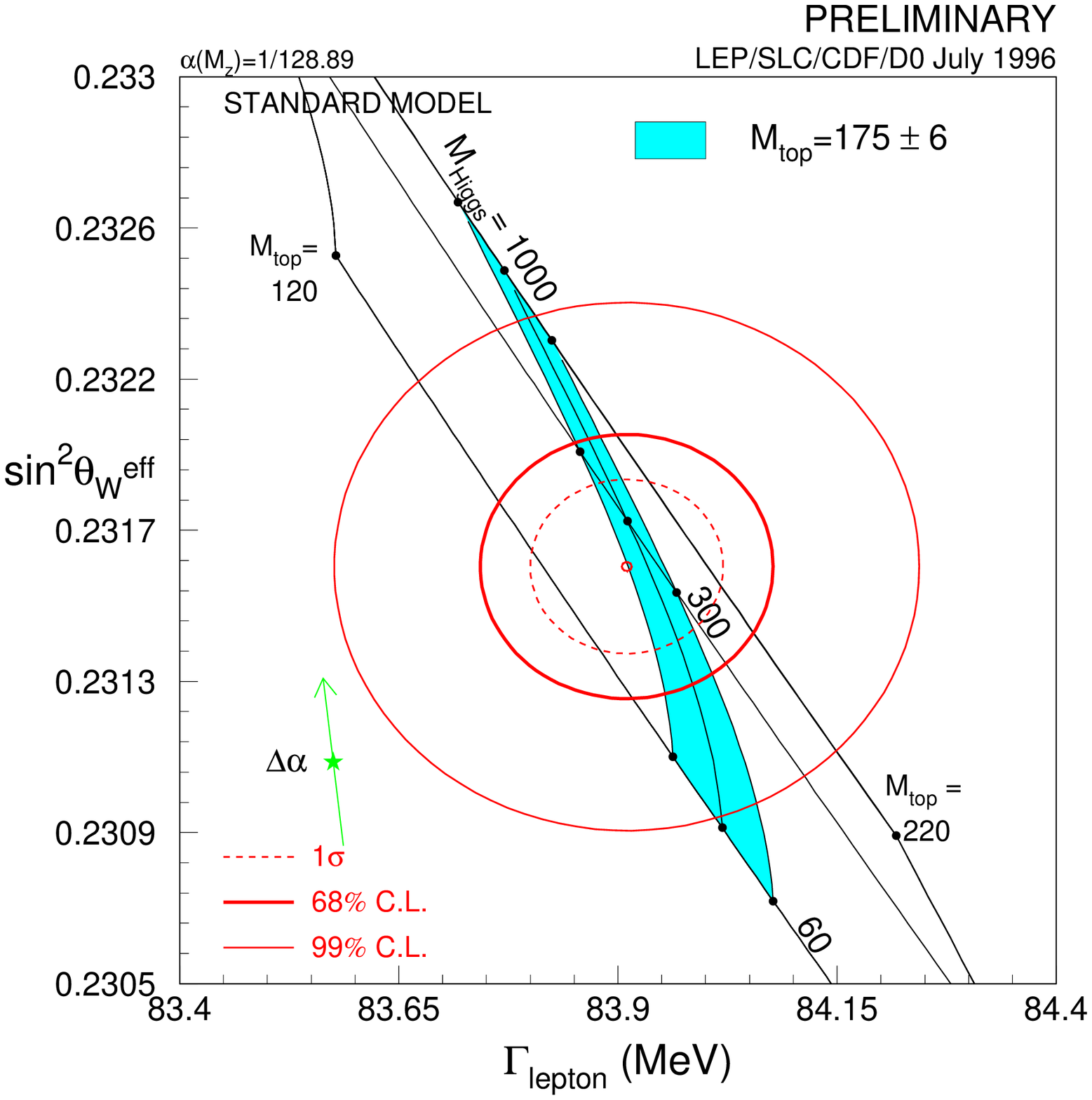}}
        \caption{The LEP/SLD measurements of $\swsqeffl$ and
                 $\Gll$ and the $\SM$ prediction.
                 The shaded area is obtained when $\Mt$ is restricted to
                 its measured mass range, $\Mt = 175 \pm 6$~\GeVcc. 
                 The star shows the predictions if among the electroweak 
                 radiative corrections only the photon vacuum polarization 
                 is included. The corresponding arrow shows the variation of 
                 this prediction if $\alpha(\MZ^2)$ is changing by one 
                 standard deviation. This variation gives an additional 
                 uncertainty on the $\SM$ prediction which is not
                 indicated in the figure. } 
        \label{fig:gll_seff}
\end{figure}

\begin{figure}[h]
        \epsfxsize = 8.0cm 
        \centerline{\epsffile{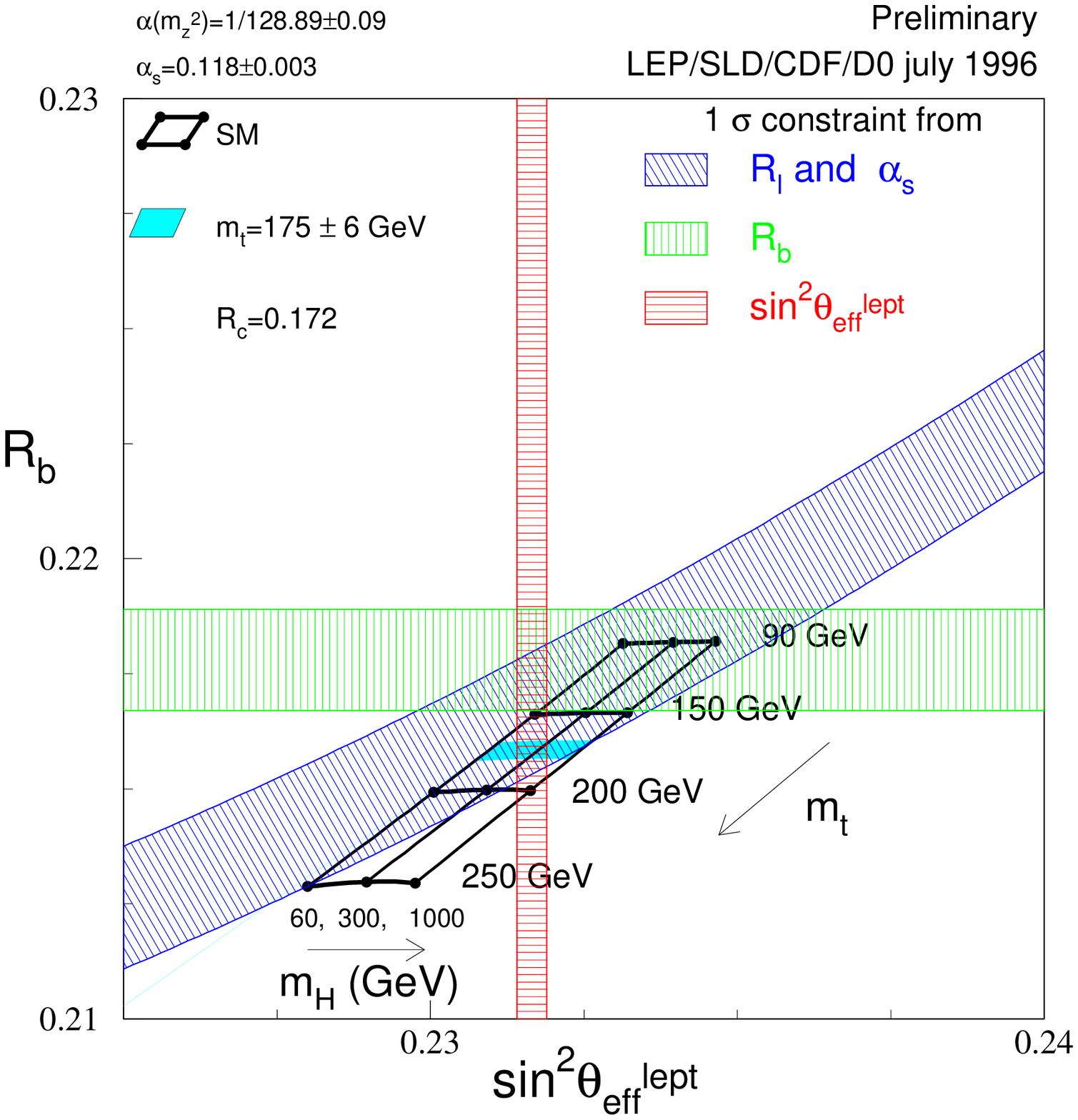}}
        \caption{The LEP/SLD measurements of $\swsqeffl$ and
                 $\Rb(\Rc=0.172)$. The grid indicates the $\SM$ 
                 prediction. Also shown is the constraint resulting from 
                 the measurement of $\RZ$, assuming 
                 $\alfmz = 0.123\pm 0.006$, as well as the 
                 $\SM$ dependence of light-quark partial widths on
                 $\swsqeffl$. } 
        \label{fig:rb_seff}
\end{figure}

The full set of observables can be fit within the framework of the
$\SM$ to up-to-date theoretical calculations~\cite{smcalc} and an
estimate of the free parameters of the model can be obtained 
along with the $\SM$ prediction for each observable. 
The accuracy of the measurements makes them sensitive to higher order
electroweak radiative corrections. The leading corrections are
due to propagator and vertex effects which introduce a dependence 
of the observables on $m_t$ (quadratically) and $\MH$ (logarithmically).
Table~\ref{tab:summary} summarizes the averages of the various
measurements from LEP (section a), from SLD (section b), 
and from electroweak
measurements from \pbarp collider and $\nu$N scattering experiments 
(section c). 
The third column tabulates the $\SM$ predictions and the last column
lists the differences between measurement and fit in units of the total
measurement error. In the $\SM$ fit the Higgs mass has been treated, for
the first time, as a free parameter. Given the multitude of
measurements, there is good agreement with the theoretical predictions. 
It seems that the only modest deviations lie within the third family: 
$\Afbzt$, which 2.3~standard deviations high, $\Afbzb$, which is 
1.8~standard deviations low, 
and $\Rb$ which has come down considerably from the earlier
measurements but is still high by 1.8~standard deviations. 
Figures~\ref{fig:gll_seff} and \ref{fig:rb_seff} give an overall
picture of the comparison with the $\SM$ in the leptonic and hadronic
sector, respectively. 
Figure~\ref{fig:gll_seff} shows a comparison with the $\SM$ of $\Gll$ 
from LEP, and $\swsqeffl$ from asymmetries measured at LEP and SLD. 
Good agreement with
the $\SM$ prediction is observed. The star indicates the prediction if
among the electroweak radiative corrections only the photon vacuum
polarization is included, showing evidence that the data is truly 
sensitive to electroweak corrections.
The length of the arrow indicates the error on $\alpha(\MZ^2)$, which is 
as large as the error on $\swsqeffl$ from LEP and SLD
combined~\cite{alpha}. 

In Fig.~\ref{fig:rb_seff} the fitted result for $\Rb$ with $\Rc$ fixed to
its $\SM$ value is plotted versus $\swsqeffl$. If one assumes the
$\SM$ dependence of the partial widths on $\swsqeffl$ for the
light quarks and the $c$ quark, and takes $\alfmz=0.123\pm 0.006$, $\Rl$
imposes a constraint on the two variables, shown as the diagonal band. 
Good agreement is seen among these three experimentally independent
measurements, showing the consistency of the LEP data.

%
%
%
%

\begin{table*}[t]
\begin{center}
{\footnotesize 
\begin{tabular}{||c||c|c|c||}
\hline
         & LEP &  LEP + SLD  &  LEP + SLD                   \\
         &     &             & + $\pbarp$ and $\nu$N data   \\
\hline
\hline
$\Mt$\hfill(\GeV/c$^2$) & $171\;\pm8\;^{+17}_{-19}$
                        & $177\;^{+7}_{-8}\;^{+17}_{-19}$
                        & $177\pm 7\;^{+16}_{-19}$          \\
$\alfmz$                & $0.122 \pm 0.003\;\pm 0.002$
                        & $0.121 \pm 0.003\;\pm 0.002$
                        & $0.121 \pm 0.003\;\pm 0.002$        \\
\hline
\hline
$\swsqeffl$             & $0.23209 \pm 0.00024\;^{+0.00007}_{-0.00016}$
                        & $0.23179 \pm 0.00022\;^{+0.00006}_{-0.00013}$
                        & $0.23179 \pm 0.00020\;^{+0.00006}_{-0.00014}$ \\
$\swsq$                 & $0.2247  \pm 0.0009\;^{+0.0003}_{-0.0002}$
                        & $0.2238  \pm 0.0008\;^{+0.0004}_{-0.0002}$
                        & $0.2238  \pm 0.0008\;^{+0.0003}_{-0.0002}$     \\
$\MW$\hfill(\GeV/c$^2$) & $80.292  \pm 0.048\;^{+0.010}_{-0.018}$
                        & $80.337  \pm 0.041\;^{+0.010}_{-0.021}$
                        & $80.338  \pm 0.040\;^{+0.009}_{-0.018}$        \\
\hline
\end{tabular}
}
\end{center}
\caption[]{Results of fits to the three sets of electroweak precision data, as
           summarized in Table~\ref{tab:summary}, for $\Mt$ and $\alfmz$. 
           The central values and the first errors quoted refer to 
           $\MH=300$~\GeV/c$^2$. 
           The second errors correspond to the variation of the central 
           value when varying $\MH$ in the interval 
           $60 < \MH < 1000$~GeV/c$^2$. 
           The bottom part of the table lists derived results for 
           $\swsqeffl$, $\swsq$ and $\MW$. }
\label{tab:mt_fit}
\end{table*}

\subsection{Predictive Power of the Standard Model }

Having shown the consistency of all the measurements with the $\SM$, it is
justified to combine them to determine 
the free parameters of the model. A beautiful precedent has been the 
prediction of the top quark mass. The top quark was discovered~\cite{mt}
in the mass region right were it was predicted to be.
Table~\ref{tab:mt_fit} shows the constraints on two free parameters of
the $\SM$, $\Mt$ and $\alfmz$, when fitting the measurements 
to $\SM$ calculations. 
No external constraint on $\alfmz$ has been imposed.
The three columns present the results corresponding to the data sets as 
listed in Table~\ref{tab:summary} sections a,~b and~c, respectively. 
The central values and the first errors quoted refer to
$\MH=300$~\GeV/c$^2$. 
The second errors correspond to the variation of the central value when 
varying $\MH$ in the interval $60 < \MH < 1000$~GeV/c$^2$. 
The bottom part of the table lists derived results for $\swsqeffl$, 
$\swsq$ and $\MW$.
The first error includes the uncertainty on the fine structure constant 
$\alpha(\MZ^2)=1/(128.896\pm 0.090)$. This large
uncertainty~\cite{alpha} is becoming
a limiting factor in the predictive power of the $\SM$. 
It causes an uncertainty of 0.00023 on the prediction of
$\swsqeffl$, an uncertainty as large as the current experimental
uncertainty, and an uncertainty of 4~\GeV/c$^2$ on $\Mt$. 
Theoretical uncertainties due to missing higher order
corrections, are neglected for the results presented in 
Tables~\ref{tab:mt_fit} and \ref{tab:mh_fit}. 
They are estimated~\cite{precision_calc} to be 
less than 1~\GeV/c$^2$ on $\Mt$, less than 0.001 on $\alfmz$ and 0.1 
on $\log(\MH)$.
Although the theoretical error on $\log(\MH)$ is still smaller than the
experimental error, it is significantly larger than the theoretical
error on $\Mt$ or $\alfmz$. Increased precision in both the fine
structure constant and the theoretical calculations is clearly warranted.

\begin{figure}[t]
    \epsfxsize = 7.0cm
    \centerline{\epsffile{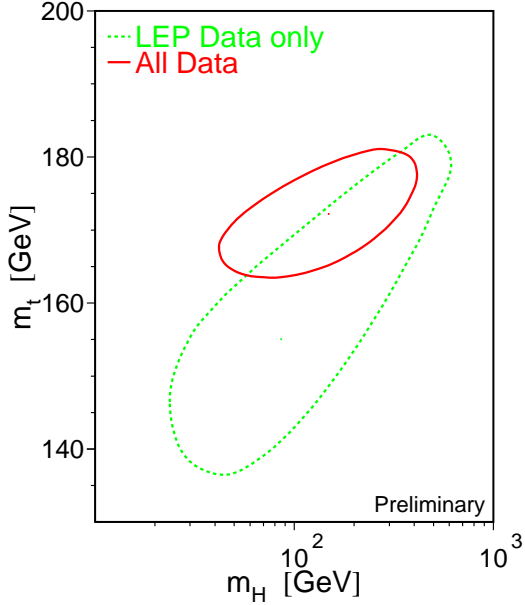}}
\caption{68\% confidence level contours in $\Mt$ and $\MH$ when using
         as constraint LEP data only (dashed line) and the world's data 
         (solid line). }
\label{fig:mt_mh}
\end{figure}

The fitted value of $\Mt$ is in excellent agreement with the measured   
top mass of $\Mt=175  \pm 6$ \GeV/c$^2$~\cite{mt}. 
Note that the precision of the direct top mass measurement has 
(finally) surpassed the indirect measurement. 
In the determination of the central value of $\Mt$, however, 
the mass of the Higgs boson has been fixed to 300~GeV/c$^2$. 
Since there is a strong correlation between the top and Higgs mass it 
should be possible to constrain $\MH$, given the Tevatron direct 
measurements of $\Mt$. 
The result of the fit is shown in Table~\ref{tab:mh_fit} and 
Fig.~\ref{fig:mt_mh}. The combination of the world's data starts to
constrain the Higgs mass and prefers a value of 
$\MH = 149^{+148}_{-82}$~GeV/c$^2$. The correlation between $\MH$ and
$\Mt$ is apparent. It should be noted that the correlation would 
even be larger if the $\Rb$ measurement is not used, as $\Rb$ is
insensitive to $\MH$.  
It should be pointed out that the central value of the preferred
Higgs mass, with the corresponding error, can vary dramatically if one
of the results is excluded from the fit. The overall constraint on 
$\MH$ is therefore still rather weak. 
The implications of these results on new physics are discussed
in~\cite{langacker}.

\begin{table}[h]
  \begin{center}
    \begin{tabular}{||c||c|c||}
\hline
                   & LEP                   &  LEP+SLD+$\pbarp$   \\
                   &                       &  +$\nu$N data+$\Mt$  \\
\hline
\hline
$\Mt$ [\GeV/c$^2$] &$155^{+18}_{-13}$      &$172 \pm 6$             \\
$\MH$ [\GeV/c$^2$] &$86^{+202}_{-51}$      &$149^{+148}_{-82}$      \\
$\log(\MH)$        &$1.93^{+0.52}_{-0.39}$ &$2.17^{+0.30}_{-0.35}$  \\
$\alfmz$           &$0.121 \pm 0.003$      &$0.120 \pm 0.003$       \\
\hline
\end{tabular}
\caption{Results for parameters in the $\SM$ from fits to LEP data 
         alone and to all data 
         including the Tevatron top quark mass determination.  } 
\label{tab:mh_fit}
\end{center}
\end{table}

\subsection{More on $W$ Properties }

\begin{figure}[h]
    \epsfxsize = 6.0cm
    \centerline{\epsffile[50 150 525 675]{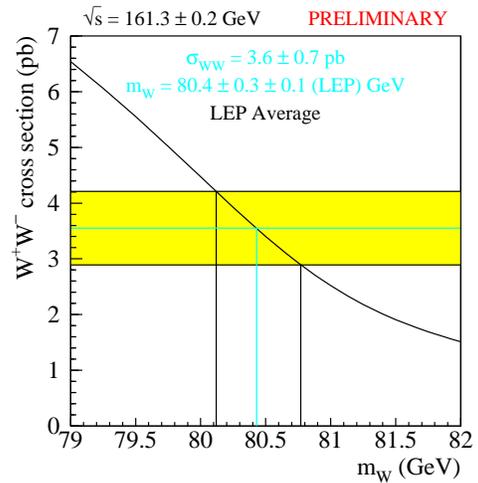}} 
\caption[]{$W$ mass from the LEP average of the measurements of threshold 
           $W$ pair production cross section at 
           ${\sqrt s} = 161.3 \pm 0.2$~GeV. }
\label{fig:wpair}
\end{figure}

Recently the LEP center of mass energy has crossed the $W$ pair
production threshold, allowing for a direct measurement of $W$ boson
properties at LEP complementing the measurements at $\pbarp$ colliders. 
One of the more interesting measurements is the $W$ mass
measurement. Given the strong sensitivity of the $WW$ production
threshold to $\MW$ a good precision is obtained with relatively few
events by measuring the total production cross section at threshold. 
Given the nature of this measurement, the dominant uncertainties are 
obviously those on the luminosity and center of mass energy. All four
LEP experiments have an initial measurement of the production cross
section at ${\sqrt s}=161.3 \pm 0.2$~GeV, listed in 
Table~\ref{tab:wpair}, resulting in a measurement of the $W$ mass of 
$\MW = 80.4 \pm 0.3 \pm 0.1$~\GeV/c$^2$~(see 
Fig.~\ref{fig:wpair})~\cite{lep2_mw}.

\begin{table}[h]
\begin{center}
\begin{tabular}{|l|c|} \hline
    ALEPH           & $4.9^{+1.9}_{-1.6}$ pb  \\
    DELPHI          & $3.5^{+1.5}_{-1.3}$ pb  \\
    L3              & $2.9^{+1.3}_{-1.1}$ pb  \\
    OPAL            & $3.9^{+1.8}_{-1.4}$ pb  \\
\hline
    LEP average     & $3.6 \pm 0.7$ pb        \\ 
\hline
\end{tabular}
\caption[]{LEP measurements of the $W$ pair production cross section 
           at ${\sqrt s} = 161.3 \pm 0.2$~GeV. }
\label{tab:wpair}
\end{center}
\end{table}

The $\SM$ process of $W$-pair production is characterized 
by large cancellations between the $s$ and $t$ channel 
production processes. The contributions from the $t$ channel
diagrams by themselves violate unitarity. The measurements of the 
pair production cross section are therefore a beautiful demonstration of
the gauge cancellations in the $\SM$, as demonstrated already with the
study of $W$ pairs produced at the $\pbarp$ colliders~\cite{my_snowmass}. 
The direct production of $W$ bosons now also allows for a direct
measurement of its magnetic dipole and electric quadrupole moment at LEP, 
two quantities on which stringent limits already exist from the 
$\pbarp$ experiments~\cite{wg_d0}.

\section{Conclusions}

It has been unprecedented that an anticipated quark was discovered with
a mass exactly within the range predicted from loop corrections within 
a theoretical framework. 
This is a remarkable feat for experimentalists and theorists alike 
and attests to the enormous success of the $\SM$. 
Even though many measurements are now being carried out with excruciating 
precision, the $\SM$ shows no signs of
giving up its claim of being the description of the fundamental
interactions as we know them. 
The large deviations that existed in the $\Rb$ and $\Rc$ measurements 
have greatly diminished. 

The $\SM$, though, is incomplete. 
Given its inherent shortcomings one gets the feeling, 
looking back at for example Fig.~\ref{fig:gll_seff} and
Table~\ref{tab:summary}, that in some sense the agreement with 
the $\SM$ predictions is too good. 
With the new data from LEP~2, SLD and the Tevatron, 
and with the planned upgrades of the accelerators as well as the
experiments, the projected uncertainties~\cite{snowmass_ewk} 
on some fundamental parameters should provide the 
tools to take another ever more critical look at the $\SM$, 
without any theoretical prejudice.

\section{Acknowledgements}
My special thanks go to all the members of the LEP Electroweak
Working Group who, as a matter of fact, provided me with a large 
fraction of the results which I presented. 
I am especially grateful to Bob Clare, 
Ties Behnke, Paul Grannis, Hugh Montgomery, Pete Renton, Dong Su 
and Roberto Tenchini who have been most cooperative!


\begin{thebibliography}{99}
\bibitem{angle} 
        The angle $\vartheta$ is the polar angle of the outgoing fermion 
        with respect to the incoming electron or proton beam.
        Pseudorapidity is defined as $\eta = -\ln\tan{\vartheta\over 2}$
\bibitem{lineshape} 
        That line shapes contain mass information is a very 
        general observation. It even holds for humans.
\end{thebibliography}
\end{document}